\documentstyle[12pt]{article}

\def\Journal#1#2#3#4{{#1}{\bf #2}, #3 (#4)}
\def\NPB{{\em Nucl.Phys.} B}
\def\PLB{{\em Phys. Lett.}B}
\def\PRD{{\em Phys.Rev.}D}
\def\PR{\em Phys.Reports}
\def\AD{\em Adv. Theor. Math. Phys.}
\def\JMP{\em Jour.Math.Phys.}
\def\CMP{\em Comm.Math.Phys.}

\def\ts{\textstyle}
\def\al{\alpha}

\def\be{\begin{equation}}
\def\ee{\end{equation}}
\def\bea{\begin{eqnarray}}
\def\eea{\end{eqnarray}}
\def\half{{1\over 2}}
\def\hhalf{{\textstyle{1\over 2}}}
\def\H{{\cal H}}
\def\F{{\cal F}}
\def\psbar{\overline{\psi}}
\def\zd{{(z - \zeta)}}

\def\dt{\cdot}
\def\bc#1,#2.{\left({#1\atop #2}\right)}
\def\X{{\cal X}}
\def\bdot{{\cdot\atop b}}
\def\t{\theta}
\def\s{\sigma}

\def\p{{\partial}}

\def\thalf{{\textstyle{3\over 2}}}
\def\nox{{\scriptstyle{\times \atop \times}}}
\def\sqr#1#2{{\vbox{\hrule height.#2pt\hbox{\vrule width
.#2pt height#1pt \kern#1pt\vrule width.#2pt}\hrule height.#2pt}}}
\def\Box{\mathchoice\sqr64\sqr64\sqr{4.2}3\sqr33}
\def\N{{\nabla}}
\def\e{\epsilon}

\begin{document}
\thispagestyle{empty}
\begin{flushright}
hep-th/0201209
\end{flushright}
\baselineskip=16pt

\vspace{.5in}
{\Large
\begin{center}
{TASI Lectures on Perturbative String Theory and Ramond-Ramond Flux}
\end{center}}

\begin{center}
L. Dolan\\ \emph {Department of Physics, University of North Carolina,\\
Chapel Hill, North Carolina 27599-3255}
\end{center}
\vspace{1in}



\begin{center}
\textbf{Abstract}
\end{center}
\begin{quotation} 
\noindent These lectures provide
an introduction to perturbative string theory
and its construction on spaces with background Ramond flux.
Traditional covariant quantization of the string and 
its connection with vertex operators and conformal
invariance of the worldsheet theory are reviewed.
A supersymmetric covariant quantization of the superstring 
in six and ten spacetime dimensions is discussed. Correlation functions
are computed with these variables. Applications to
strings in anti-de Sitter backgrounds with Ramond flux
are analyzed. 
Based on lectures presented at the Theoretical Advanced Study Institute
TASI 2001, June 3-29, 2001.
\end{quotation}

\newpage

\tableofcontents

\newpage

\section{Introduction}

Recently, the promise of solving strong coupling Yang-Mills theory
by considering the Type IIB superstring on anti de Sitter space ($AdS$)
via the $AdS/CFT$ correspondence has led to a renewed interest in
perturbative string theory and its formulation on background 
curved spaces. The first lecture reviews the derivation of the
physical spectrum and scattering amplitudes 
in the old covariant quantization for
open and closed bosonic string theory, with attention given to
the structures that will require modification when the
background spacetime is curved. The second lecture
reviews various worldsheet formulations for the superstring,
including the Berkovits-Vafa-Witten variables which
provide a manifestly supersymmetric and covariant quantization
in six dimensions. 
In the third lecture, these worldsheet fields 
are used to solve the string constraints
on the vertex operators for the Type IIB superstring
on $AdS_3\times S^3\times K3$ with background Ramond flux. 
A short section on computing correlation functions
using these fields has been added in 4.3.

\section{Old Covariant Quantization}

We review
the traditional quantization~{\cite{physrepone}-\cite{gsw}} 
of the open and closed bosonic strings and point to the steps that 
need generalization to accommodate strings with background
Ramond fields. 

\subsection{Open Bosonic String}

We introduce the Fubini-Veneziano fields:
\be X^\mu (z) = q^\mu - ip^\mu\ln z + i\sum_{n\ne 0}{a_n^\mu\over n} z^{-n}
\label{eq:fv}\ee
where 
\be [a_m^\mu , a_n^\nu] = \eta^{\mu\nu} m\delta_{m,-n} ;\quad
[q^\mu , p^\nu] = i\eta^{\mu\nu} ; \quad [q^\mu, a_n^\nu] = 0, n\ne 0;\quad
n\in{\bf Z}\,.\label{eq:qc}\ee
These commutators are the Lorentz covariant quantization conditions.
Here $p^\mu\equiv a_0^\mu$.
The fields $X^\mu (z)$ are restricted to appear in an exponential or as a 
derivative, since they do not exist rigorously as quantum fields which have 
a well-defined scaling dimension. The metric is space-like,
$\eta_{\mu\nu} = {\rm{diag}} (-1, 1, ... 1)$, $\mu = 0, 1, ..., d-1$; and
the $a_n^\mu$ satisfy the hermiticity relations $a_n^{\mu\dagger} = 
a^\mu_{-n}$. In flat spacetime, momentum is conserved and it is
convenient when quantizing to use a basis of momentum eigenstates,
$|k\rangle = e^{ik\cdot q} |0\rangle,\,  p^\mu |k\rangle = k |k\rangle,$
to represent $[q^\mu , p^\nu] = i\eta^{\mu\nu}$,
where the vacuum state $|0\rangle$ satisfies
$a_n^\mu|0\rangle = 0, n\ge 0$.  When the spacetime contains $AdS$, 
the isometry group no longer contains translations, so we lose
momentum conservation and will just work in position space.

A {\it string} is a one-dimensional object that moves through spacetime
and is governed by an action
that describes the area of the worldsheet. Its trajectory
$x^\mu (\sigma,\tau)$ describes
the position of the string in space and time. $\tau$ is the evolution parameter
$-\infty <\tau <\infty$, and $\sigma$ is the spatial coordinate labelling 
points along the string; $0\le\sigma\le\pi$ for the open string, which is
topologically an infinite strip; and $0\le\sigma\le2\pi$ for the closed string
which is topologically an infinite cylinder
with periodicity condition $x^\mu (\sigma,\tau) = x^\mu (\sigma + 2\pi, \tau)$.
$g^{\alpha\beta}(\sigma,\tau)$ is
the two-dimensional metric, $\alpha,\beta = 0,1$.
The action for the bosonic string is
\bea S_2 &= -{1\over 4\pi\alpha '} \int d\sigma d\tau [\, \sqrt{|g|} 
g^{\alpha\beta}
\,\partial_\alpha x^\mu\partial_\beta\,x^\nu \,G_{\mu\nu}(x) \cr
&\hskip50pt +\epsilon^{\al\beta} \partial_\alpha x^\mu\partial_\beta\,x^\nu
\, B_{\mu\nu}(x)\cr
&\hskip30pt + \al' \sqrt{|g|} \,R \,\phi (x)\,]\,. 
\label{eq:wsa}\eea
The Regge slope $\alpha '$ has dimensions of length squared $[L]^2$; 
the string tension is defined as 
$T = {1\over {2\pi\alpha '}}$. The low energy limit
of the string theory is an effective point field theory. This corresponds
to the leading term in an expansion
in the external momentum times $\sqrt{\alpha '}$. 
In the zero Regge slope limit,i.e. the infinite string tension limit, 
the interactions of the vectors and
tensors are those of Yang-Mills bosons and gravitons. 
In the other limit, the zero tension limit, 
one expects an infinite number of massless particles.
It is conjectured for strings on $AdS$ that both limits may be simple field 
theories, since interacting massless particles of spin higher than $2$
appear to be consistent in an Einstein spacetime, whereas they are not in flat
space. 

For general $G_{\mu\nu} (x)$,
from a $2d$ point of view,
$S_2$ is a nontrivial interacting field theory: a conformally invariant
non-linear sigma model. To quantize in flat spacetime,
we choose $G_{\mu\nu} (x) = \eta_{\mu\nu}$, and the other background
fields to vanish: the two form field potential $B_{\mu\nu}$ 
and the dilaton $\phi$. This choice reduces $S_2$ to 
a free worldsheet theory. As a result, the correlation 
functions even at tree level in the string loop expansion are 
exact in $\alpha'$. The Type IIB superstring 
on $AdS_5\times S^5$ requires additional background fields
besides the curved metric $G_{\mu\nu} (x)$. In this latter case,
$S_2$ is a nontrivival worldsheet theory where the
string tree level amplitudes will appear as an
expansion in $\alpha'$.

The string has two 
parameters: the length scale ${\sqrt{2\alpha '}}$ 
and the dimensionless string coupling constant 
$g$. These are both related to the dilaton vacuum
expectation value $\langle 0|\phi |0\rangle$, and thus the string 
has no free dimensionless (nor dimensionful) 
parameters. The gravitational coupling is $\kappa\equiv\sqrt{8\pi G}$,
Newton's constant is $G = {\hbar  c\over m^2_{PLANCK}}$;
the Planck mass is $m_{\rm PLANCK} = ({\hbar c \over G})^{1\over 2} \sim
2.2\times 10^{-5} {\rm gm} \sim 1.2\times 10^{19} GeV$; the Planck length is
$l_{\rm PLANCK} = ({\hbar G \over c^3})^{1\over 2} \sim 1.6\times 10^{-33}$cm.
The Planck time is $t_{\rm PLANCK} = ({\hbar G \over c^5})^{1\over 2}
\sim 10^{-43}$ sec.\hfil\break
At these scales, the effects of stringiness will be important, whereas
at larger distance scales or lower energies, an ordinary point
quantum field theory QFT can be used as an
effective theory. In flat space and for
$\langle 0| \phi |0\rangle = 0 $, 
from identification of the graviton vertices we find
$\kappa = {\half} g \sqrt {2\alpha'}$, 
and from the Yang-Mill vertices, $g_{ym} = g$.
Then
$\alpha ' \sim {1\over {m_{PLANCK}^2}}$ i.e.
$\alpha ' \sim\kappa ^2$, specifically $\alpha' = 2\kappa^2 / g^2 =
16\pi G / g^2$.
Thus the value of the universal 
Regge slope parameter $\alpha'$ is given in terms of 
Newton's constant $G$ and the Yang-Mills coupling $g$. In particular since $g$
is of order 1, $\alpha'$ is of order the Planck length (squared).
We see that the scale of the entire unified string 
theory is set by the Planck mass. 
This scale does not appear to be associated with the secret of any symmetry
breaking, as does the scale $\sqrt{\lambda /3} \,a = \sqrt{-2m^2}\equiv m_H$
given by the Higgs mass $m_H \sim 250\, Gev $. Discussions of 
the gauge hierarchy problem, {\em i.e.}
why is $ m / m_{\rm PLANCK}$ so small,
what sets the ratio of $m$ to $m_{\rm PLANCK}$?
and what is the origin of mass, {\em i.e.} how is $m\ne 0$?
are beyond the scope of these lectures. 

The worldsheet action $S_2$ has
two-dimensional general coordinate invariance
$g^{\alpha\beta}\rightarrow 
\partial_\gamma\Lambda^\alpha\partial_\delta\Lambda^\beta g^{\gamma\delta}
\sim g^{\alpha\beta} + f^\gamma\partial_\gamma g^{\alpha\beta} +
\partial_\gamma f^\alpha g^{\beta\gamma} + 
\partial_\gamma f^\beta g^{\alpha\gamma}$     
and $x^\mu\rightarrow x^\mu + f^\alpha\partial_\alpha x^\mu$; and local
Weyl rescaling
$g^{\alpha\beta}\rightarrow \Lambda (\sigma,\tau) g^{\alpha\beta}$ 
and $x^\mu\rightarrow x^\mu$.
These symmetries allow one to make the {\it covariant gauge choice}
$g^{\alpha\beta} = \eta^{\alpha\beta} = {\rm diag} (-1,1)$.
This results in the remnant constraint equations
$(\partial_\sigma x^\mu\pm\partial_\tau x^\mu)^2 = 0\,$
which are equivalent to
$\partial_\sigma x\cdot\partial_\tau x = 0$ and
$\partial_\sigma x\cdot\partial_\sigma x +
\partial_\tau x\cdot\partial_\tau x = 0.$

The equations of motion are
$\partial^2_\sigma x^\mu - \partial^2_\tau x^\mu
= 0 .$
The open string boundary conditions are
$\partial_\sigma x^\mu (\sigma,\tau) = 0 \,{\rm at}\, 
\sigma = 0,\pi.$
The general solution of the equations of motion
with these string boundary conditions is
\bea &x^\mu (\sigma,\tau) = q^\mu + p^\mu\tau + i\sum_{n\ne 0}{a_n^\mu\over n}
e^{-in\tau} \cos n\sigma\cr
&\hskip90pt = q^\mu + p^\mu\tau +{i\over 2}\sum_{n\ne 0}{a_n^\mu\over n}
(e^{-in(\tau+\sigma)} + e^{-in(\tau-\sigma)})\cr
&\hskip20pt = {1\over 2}X^\mu (e^{i(\tau+\sigma)}) + {1\over 2}X^\mu 
(e^{i(\tau-\sigma)})\,. \label{eq:sol}\eea
In covariant gauge, we implement the constraint equations by observing
that
\be (\partial_\sigma x^\mu\pm\partial_\tau x^\mu)^2 = 
z^2 L(z)\,,\qquad z = e^{i(\tau\pm\sigma)}\label{eq:ceq}\ee
where 
\be L(z) = {\ts{1\over2}}:a(z)\cdot a(z): 
= \sum_n L_n z^{-n-2}\label{eq:vcur}\ee
and $a^\mu (z)\equiv i{dX^\mu (z)\over dz} = \sum_n a_n z^{-n-1}$.
So the expectation value of the constraints vanish 
(as $\hbar \rightarrow
0$) for the {\em physical state conditions} in covariant gauge
in the old covariant quantization:
\be L_0|\psi\rangle = |\psi\rangle\label{eq:onecon}\ee
\be L_n|\psi\rangle = 0 \quad {\rm for}\, n>0\,.\label{eq:twocon}\ee

Furthermore (\ref{eq:onecon}) is the mass shell condition $p^2 = -m^2$,
where $\half m^2\equiv N - 1$ so that $L_0 = {1\over 2} p^2 + N = 1$.
Here $N = \sum_{n=1}^\infty a_{-n}\cdot a_n$.
That is to say a physical state $|\psi\rangle$ has momentum $k$ which takes
on a specific value corresponding to the $N^{th}$ excited level,
$\alpha ' k^2 = 1 - N$.
The physical state conditions (\ref{eq:onecon},\ref{eq:twocon})
can be shown to eliminate ghosts,
{\it i.e.} negative norm states when $d = 26$.
This result is known as the No-Ghost theorem~\cite{gt,b}.

If instead, we had considered light-cone gauge, we would have
observed that in the constraints 
$(\partial_\sigma x^\mu\pm\partial_\tau x^\mu)^2 = 0$
there is still residual gauge invariance 
to choose the light-cone gauge where $x^+ (\sigma,\tau) = q^+ + p^+\tau$,
and the equations of motion are
$(\partial^2_\sigma - \partial^2_\tau )\, x^i
= 0\,,$ $1\le i\le d-2$.

\subsection{Locality of Worldsheet Fields}

A conformal field theory $\H$ is a Hilbert space of states $H$,
such as the space of finite occupation number states in a Fock space,
together with
a set of vertex operators $V(\psi,z)$, 
i.e. conformal fields which are in one to 
one correspondence with the states $\psi\in\F (H)$, where $\F (H)$ is a
dense subspace of the Hilbert space $H$ of states.
\be \H = (H, \{V(\psi,z): \psi\in\F (H)\})\,.\ee 
The conformal field theory requires that the vertex operators $V(\psi,z)$
form a system of mutually local fields, where  
$\lim_{z\rightarrow 0} V(\psi,z) |0\rangle = \psi$ for
each field. That is to say
the conformal fields $V(\psi,z)$ acting on the vacuum create asymptotic ``in''
states $\psi = V(\psi,0) |0\rangle$ with conformal weight $h_\psi$,
$L_0 \psi = h_\psi \psi$ (recall that $z=0$ is $i\tau = t = -\infty$) on the
cylinder. 
There is a one to one correspondence between the fields and the states
in the Hilbert space they create at $z=0$.
Locality implies the s-t duality relation
$V(\psi ,z)V(\phi ,\zeta)=V(V(\psi ,z-\zeta)\phi ,\zeta )$
which provides a precise version of the operator product expansion~\cite{dgm}.
In particular locality requires
\be V(\psi ,z) V(\phi ,\zeta) \sim V(\phi ,\zeta)  V(\psi ,z)\ee
where the left side is defined for $|z|>|\zeta|$,
the right side for $|\zeta| > |z|>$ 
and $\sim$ denotes analytic continuation.
Locality ensures well defined scattering amplitudes.
We shall also assume that the theory has a hermitian structure,
in the sense
that there is a definition of conjugation on the states, $\psi\mapsto
\psbar$, an antilinear map with
$V(\psbar ,z)=z^{-2h}V(e^{z^\ast L_1}\psi,1/z^\ast)^\dagger\,.$

\subsection{Virasoro Algebra}

One of the conformal fields is the Virasoro current
\be L(z) = V( \psi = \half a_{-1}\cdot a_{-1} |0 \rangle , z )
= \sum_n L_n z^{-n-2}\,.\ee
It satisfies the {\it operator product expansion}
\bea L(z) L(\zeta) = {c\over 2} (z-\zeta)^{-4} 
+ 2L(\zeta) (z-\zeta)^{-2} + {dL(\zeta)\over d\zeta} (z-\zeta)^{-1}
\eea
which can be reexpressed as either of 
\bea
&\hskip50pt [L_n, L(\zeta)] = 2(n+1)\zeta^n L(\zeta)
+ \zeta^{n+1} {dL(\zeta)\over d\zeta} 
+ \zeta^{n-2} {c\over 12} (n^3-n)\cr
&[L_n, L_m] = (n-m) L_{n+m} + {c\over 12} (n^3-n)\delta_{n,-m}\,.\eea
Vertex operators for {\em physical states} are primary fields.
In this case $\phi$ is a highest weight state for the
Virasoro algebra, or {\it primary state},
$V(\phi, \zeta)$ is a {\it primary field} and
\be L(z)V(\phi, \zeta) = 
h_\phi\zd^{-2}V(\phi, \zeta) + \zd^{-1}{dV\over d\zeta}(\phi, \zeta)\,.\ee
Therefore in covariant gauge, the vertex operators for physical states
have to satisfy~\cite{{fms},{lt}}
\be [L_n,\,V(\psi, z)]= z^{n+1}{d\over dz} V(\psi, z) 
+ (n+1) z^n V(\psi, z)\label{eq:prims}\ee
for all $n$, and
\be \lim_{z\rightarrow 0} V(\psi,z) |0> = \psi\,.\ee
This follows from
$V(z) = \sum_r V_r z^{-r-h}\,;$ so that
$[L_n, V_r ] = ( - r + n(h-1) ) V_{n+r}$. Since
$V_r |0 > = 0$ for $r>-h$, and $\psi = V(0) |0> = V_{-h} |0>$,
then $L_n \psi = [ L_n, V_{-h} ] |0> = 0$ for $n\ge 1\,,$
and physical fields are primary fields of conformal dimension $h=1$.
(Note that if (\ref{eq:prims}) holds only for $n=0,\pm 1$
then $V(\phi, \zeta)$ is a {\it quasi-primary field}.)

\subsection{Mass Spectrum and Tree Level Amplitudes}

We now consider the first few mass levels. 
For $N = 0$, there is one state $\psi = |k\rangle$ with $k^2 = 2$.
The vertex operator for this state is
\be V(k, z) = : \exp \{ik\dt X(z)\}:
= \exp \{ik\dt X_<(z)\} e^{ik\dt q} z^{k\dt p}
\exp \{ik\dt X_>(z)\} \ee
where
$X^\mu_{>\atop <}(z)= i \sum_{n{>\atop <}0} {\al^\mu_n\over
n}z^{-n}$.
Since $z^{L_0} V(k,1) z^{-L_0} = z V(k,z)$,
the four point open string tree amplitude for these tachyonic scalars is
\bea
&\hskip-50pt A_4 = 
\alpha '\int_0^1 dz \,\langle 0; -k_1|V(k_2,1) V(k_3,z)|0;
k_4\rangle \cr
&\hskip30pt = \alpha '(2\pi)^{26}\delta^{26}(k_1+k_2+k_3+k_4)
\int_0^1 dz \,z^{k_3\cdot k_4} (1-z)^{k_2\cdot k_3}\cr
&\hskip20pt = \alpha ' (2\pi)^{26}\delta^{26}(k_1+k_2+k_3+k_4)
B(-1-\half s, -1-\half t)\cr
&\hskip25pt = \alpha ' (2\pi)^{26}\delta^{26}(k_1+k_2+k_3+k_4)
B(-1-\alpha' s, -1-\alpha' t)\cr
&\hskip-1000pt 
= A(s,t)\,.\eea
Here the Mandelstam variables are $s = -(k_1 + k_2)^2\,,$ 
$t = -(k_2 + k_3)^2\,,$ and $u = -(k_1 + k_3)^2\,;$
the overall
factor of $\alpha'$ is due to the propagator.
As mentioned earlier, this string amplitude is exact in $\alpha'$,
reflecting the fact the worldsheet theory has free operator
products $a^\mu(z) a^\nu(\zeta) = \eta^{\mu\nu} (z-\zeta)^{-2}\,.$
Here $a^\mu(z)=\sum_n a_n z^{-n-1}\,.$
The three point open string tree amplitude for these tachyonic scalars is
$A_3 =\langle 0; -k_1|V(k_2,1)|0;
k_3\rangle 
= (2\pi)^{26}\delta^{26}(k_1+k_2+k_3+k_4)\,.$
The first excited level, $N = 1$, contains the vector states
$\psi = \epsilon\dt a_{-1}|k\rangle$ with $k^2 = 0$.
In order to satisfy $L_n \psi = 0$ for $n>0$, (in particular $L_1 \psi = 0$)
we must have $\epsilon\dt k = 0$.  
The vertex operator for this state is
\be V(k,\epsilon, z) = \epsilon\dt a(z) e^{ik\dt X(z)}\,.\ee
The three point open string tree amplitude for these massless vectors is
\bea 
&\hskip-180pt A_3 = \langle -k_1|\epsilon_1\dt a_1
V(k_2,\epsilon_2,1)\epsilon_3\dt a_{-1}|k_3\rangle \cr
& = (2\pi)^{26}\delta^{26}(k_1+k_2+k_3)
\, \sqrt {2\alpha '}
(\epsilon_1^\alpha\epsilon_2^\mu\epsilon_3^\lambda t_{\alpha\mu\lambda}(k_i) +
2\alpha '
(\epsilon_1\dt k_2 \epsilon_2\dt k_3
\epsilon_3\dt k_1))\,,\label{eq:bosglu}\cr
&\eea
where
\be t_{\alpha\mu\lambda}(k_i) \equiv
k_{2 \alpha} \eta_{\mu\lambda} + k_{3 \mu} \eta_{\lambda\alpha} +
k_{1 \lambda} \eta_{\alpha\mu}\ee
and
we have recovered the dependence on the dimensional parameter $\alpha'$
by dimensional analysis, {\it i.e.} $k\rightarrow \sqrt {2\alpha'}k$.
The four point open string tree amplitude for these massless vectors
has tachyon poles and is given by~\cite{schprtwo} 
\bea
&\hskip-100pt A_4 = \int_0^1 dz \,\langle -k_1|\epsilon_1\dt a_1
V(k_2,\epsilon_2,1) V(k_3,\epsilon_3,z)\epsilon_4\dt a_{-1}|k_4\rangle \cr
&=\int_0^1 dz \,z^{-\alpha's}\, 
\langle -k_1 -k_2 |\epsilon_1\dt a_1
\epsilon_2\dt a(1) e^{-k_2\dt \sum_{n<0} {\al^\mu_n\over n}z^{-n}}
e^{-k_2\dt \sum_{n>0} {\al_n\over n}z^{-n}}\cr
&\hskip 50pt\dt e^{-k_3\dt \sum_{n<0} {\al_n\over n}z^{-n}}
e^{-k_3\dt \sum_{n>0} {\al_n\over n}z^{-n}} 
\epsilon_3\dt a(z)\
\epsilon_4\dt a_{-1}|k_4\rangle \cr 
&\hskip-80pt =(2\pi)^{26}\delta^{26}(k_1+k_2+k_3+k_4)
\int_0^1 dz \,z^{-\alpha's - 2}\,(1-z)^{-\alpha't}\cr
&\hskip-220pt \dt[\epsilon_1\dt\epsilon_2 \epsilon_3\dt\epsilon_4\cr
&\hskip -100pt -\epsilon_1\dt\epsilon_2 \epsilon_4\dt k_3 \{ \epsilon_3\dt k_4
-\epsilon_3\dt k_2 z (1-z)^{-1}\}\cr
&\hskip -90pt -\epsilon_3\dt\epsilon_4 \epsilon_1\dt k_2 \{- \epsilon_2\dt k_1 
+\epsilon_2\dt k_3 z (1-z)^{-1}\}\cr
&\hskip -80pt -\epsilon_1\dt k_2\epsilon_4\dt k_3\{\epsilon_2 \dt \epsilon_3
z (1-z)^{-2} -\epsilon_2\dt k_1\epsilon_3\dt k_4\cr
&\hskip 25pt
+(\epsilon_3\dt k_4\epsilon_2\dt k_3 + \epsilon_2\dt k_1\epsilon_3\dt k_2)
z (1-z)^{-1}
-\epsilon_3\dt k_2\epsilon_2\dt k_3 z^2 (1-z)^{-2}\}]\,.\cr
&\eea
\subsection{Closed Bosonic String}

The closed string satifies the same equations of motion
$\partial^2_\sigma x^\mu - \partial^2_\tau x^\mu
= 0 $ but is toplogically a cylinder with boundary condition
$x^\mu (\sigma,\tau) = x^\mu (\sigma + 2\pi, \tau)$.
The general solution is
\bea
x^\mu (\sigma,\tau)&= q^\mu + p^\mu\tau +
{i\over 2}\sum_{n\ne 0}\{{a_n^{L\mu}\over n}
e^{-in(\tau+\sigma)} + {a_n^{R\mu}\over n} e^{-in(\tau-\sigma)}\}\cr
&= {1\over 2}X_L^\mu (e^{i(\tau+\sigma)}) + {1\over 2}X_R^\mu
(e^{i(\tau-\sigma)})\eea
where 
$X_L^\mu (z) = 
q^\mu - ip_L^\mu\ln z + i\sum_{n\ne 0}{a_n^{L\mu}\over n} z^{-n}$,
$X_R^\mu (z) = 
q^\mu - ip_R^\mu\ln z + i\sum_{n\ne 0}{a_n^{R\mu}\over n} z^{-n}$,
and $p_L = p_R = p$.
The covariant quantization conditions are 
\be [a_m^{L\mu} , a_n^{L\nu}] = [a_m^{R\mu} , a_n^{R\nu}] =
\eta^{\mu\nu} m\delta_{m,-n} ;\quad
[q^\mu , p^\nu] = {i\over 2}\eta^{\mu\nu} 
\ee
\be [q^\mu, a_n^{L\nu}] = [q^\mu, a_n^{R\nu}] = 0, \quad n\ne 0
\ee
\be 
[a_m^{L\mu} , a_n^{R\nu}] = 0;\quad a_0^{L\mu}\equiv a_0^{R\mu}\equiv p^\mu\,.
\ee
The physical state conditions are
\be
L^L_0|\psi\rangle = |\psi\rangle\, ,\quad 
L^L_n|\psi\rangle = 0 \,,\quad {\rm for}\, n>0\, ,
\ee
\be L^R_0|\psi\rangle = |\psi\rangle\, ,\quad 
L^R_n|\psi\rangle = 0 \,,\quad 
{\rm for}\, n>0\,.\ee
The mass shell condition is $p^2 = -m^2$ where $ m^2 =
N_L + N_R -2$ and $N_L = N_R$. 
$L_n^R, L_N^L$
are two commuting Virasoro
algebras, both with $c = d = 26$. In this quantization of the closed 
string in flat spacetime, the theory is seen to be a 
tensor product of left and right copies of the open string case.
One can form vertices for the closed string
by taking the tensor products of open string conformal fields for the left- and 
right movers, and using the variable $z$ for the left vertices and
$\bar z$ with the right vertices. 
Here we have defined a euclidean world sheet metric $i\tau \equiv t$, so that
$z = e^t e^{i\sigma}$, $\bar z = e^t e^{-i\sigma}$, which maps the cylinder
traced by the moving string onto the complex plane~\cite{fms}.
Time ordering is radial ordering; 
$t =$ constant hypersurfaces are circles
concentric about the origin of the $z-plane$. In fact the local
operator product relations will extend naturally to arbitrary Riemann 
surfaces with
local conformal coordinates $z$,$\bar z$.
Since for closed strings, $(L_0^L + L_0^R) |\psi\rangle = 2 |\psi\rangle$,
the tachyon $|k\rangle \equiv e^{i2k\dt q}|0\rangle$, 
$k^2 = 2$ has vertex operator
\be V(k ,z,\bar z) = :\exp \{ik\dt X_L(z)\}: :\exp \{ik\dt X_R(\bar z)\}: 
\ee
When the spacetime is not flat, the conformal fields in general
will not factor into left times right, although 
the two copies of the Virasoro algebra will remain holomorphic
(and antiholomorphic)~\cite{bvw}.

The four point closed string tree amplitude for the tachyonic scalars
in flat spacetime is again exact in $\alpha'$:
\bea
&\hskip-100pt A_4= {\alpha '\over 2\pi}
\int d^2z \,\langle -k_1|V(k_2,1,1) z\bar z V(k_3,z,\bar z)|k_4\rangle \cr
&\hskip-12pt = {\alpha '\over 2\pi} (2\pi)^{26}\delta^{26}(k_1+k_2+k_3+k_4)
\int d^2z \,|z|^{2k_3\cdot k_4} |1-z|^{2k_2\cdot k_3}\cr
&= {\alpha '\over 2\pi} (2\pi)^{26}\delta^{26}(k_1+k_2+k_3+k_4)
{\Gamma (-1-{\alpha' s}) \Gamma (-1-{\alpha' t})
\Gamma (-1-{\alpha' u})\over \Gamma (2+{\alpha' s})
\Gamma (2+{\alpha' t}) \Gamma (2+{\alpha' u})}\,\eea
which is totally symmetric under the exchange of $s,t,u$.
The massless level $N_L = N_R = 1$ contains the states
$\psi = \epsilon ^L\dt a^L_{-1} \epsilon ^R\dt a^R_{-1}
|k\rangle$ with $k^2 = 0$, and $\epsilon ^L\dt k = \epsilon ^R\dt k = 0$,
forming the spin two
graviton, the 2-form antisymmetric tensor, and the
dilaton.

The preceeding analysis of physical state conditions can be reexpressed 
in an equivalent formulation using BRST cohomology~\cite{fms}, which includes
both the ``matter'' sector described above and a BRST ``ghost'' sector
with equal and opposite central charge. 
The worldsheet variables used to 
formulate strings in curved spacetime with Ramond flux, 
do not exhibit such a `matter times ghost'' 
factorization~\cite{bv,bvw,berkthree},
although they still define 
physical state conditions as a version of cohomology.

\section{Various Formulations of Superstring Worldsheet Fields}

\subsection{Ramond-Neveu-Schwarz ($RNS$)}

This a Lorentz covariant but not {\it manifestly} supersymmetric
quantization. The Neveu-Schwarz ($NS$) fields 
$b^\mu(z) = \sum_s {b_s^\mu} z^{-s-\hhalf}$ have 
$\{b_r^\mu ,b_{s}^\nu\} = \eta^{\mu\nu} \delta_{r,-s};$ where
$r,s\in {\bf Z} + \hhalf\,$ and
$b_s^{\mu\dagger} = b_{-s}^\mu\,. $ They are used to construct
the super Virasoro generators
$G(z) = a(z)\dt b(z)$ and
$L(z) = \hhalf :a(z)\cdot a(z): + \hhalf:{db(z)\over dz}\cdot b(z):$
which satisfy the $N=1$ super Virasoro algebra
\bea
&[L_m, L_n] = (m-n) L_{n+m} + {c\over 12} (m^3 - m)\delta_{m,-n}\cr
&\hskip-53pt 
\{G_r, G_s\} = 2 L_{r+s} + {c\over 3} (r^2 - {1\over 4})\delta_{r,-s}\cr
&\hskip-90pt [L_n, G_s] = ({n\over 2}- s) G_{n+s}\,.\eea
The No-Ghost theorem selects $c={\textstyle{3\over 2}}d = 15$.
The {\it physical state} conditions in the ${\cal F}_2$-picture are
$L_0|\psi\rangle = \half |\psi\rangle\,,$
$L_n|\psi\rangle = 0\,$ for $n>0,$
$G_s|\psi\rangle = 0\,$ for $s>0\,.$
In a superconformal field theory the states are in one-to-one correspondence
with a conformal superfield $V(\psi, z,\vartheta) = V_0(\psi, z) +
\vartheta V_1(\psi, z)$, where $\vartheta$ is a fermionic coordinate,
the supersymmetry partner of $z$. $V_0(\psi, z)$ and $V_1(\psi, z)$ 
are called the lower and upper components of the superfield respectively.
\be
\lim_{z\rightarrow 0} V_0(\psi,z) |0\rangle = |\psi\rangle\,,\quad
\lim_{z\rightarrow 0} V_1(\psi,z) |0\rangle = G_{-\half}|\psi\rangle
\,.\ee
In covariant gauge, the vertex operators for physical states
have to satisfy
\bea
&[L_n,\,V_0(\psi, z)] = z^{n+1}{d\over dz} V_0(\psi, z)
+ h (n+1) z^n V_0(\psi, z)\cr
&\hskip28pt [L_n,\,V_1(\psi, z)] = z^{n+1}{d\over dz} V_1(\psi, z)
+ (h+\hhalf) (n+1) z^n V_1(\psi, z)\cr
&\hskip-112pt
\{G_s,\,V_0(\psi, z)\}_{\pm} = z^{s+\half} V_1(\psi, z)\cr
&[G_s,\,V_1(\psi, z)]_{\mp} = z^{s+\half} {dV_0(\psi, z)\over dz} 
+ 2h (s+\half) z^{s-\half} V_0(\psi, z)\eea
for all $n$ and $s$, for $h = \half$.
At level $N = 0$, there is one state $\psi = |k\rangle$ with $k^2 = 1$.
Its vertex operator has components
\be V_0(k, z) = : \exp \{ik\dt X(z)\}:\, ,\quad
V_1(k, z) = \sqrt {2\alpha'}\,  k\dt b(z) : \exp \{ik\dt X(z)\}:\, .\ee
Unlike the bosonic case, 
the three point amplitude for the Neveu-Schwarz tachyon vanishes:
$A_3 = \langle -k_1 | V_1 (k_2, 1) |k_3\rangle = 0\,.$

The massless vector is at level $N = \half$,
$\psi = \epsilon\dt b_{-\half}|k\rangle$ with $k^2 = 0$.
To satisfy $L_n \psi = 0$ for $n>0$, $G_s \psi = 0$ for $s>0$
(in particular $G_{\half} \psi = 0$)
we must have $\epsilon\dt k = 0$.
Its vertex operator has components
\bea
&V_0(k,\epsilon, z) = \epsilon\dt b(z) e^{ik\dt X(z)}\,,\cr
&V_1(k,\epsilon, z) = \left\{
\sqrt {2\alpha '}k\dt b(z)\,\epsilon\dt b(z)
+ \epsilon\dt a(z) \right\}  
\exp \{ik\dt X(z)\}\,.\eea
For this state,
$G_{-\half}\psi = (k\dt b_{-\half}\epsilon\dt b_{-\half} + \epsilon\dt a_{-1})
|k\rangle$.
The three point open string tree amplitude for these massless vectors is
\bea
&A_3 = \langle -k_1|\epsilon_1\dt b_\half
V_1(k_2,\epsilon_2,1)\epsilon_3\dt b_{-\half}|k_3\rangle \cr
&= (2\pi)^{10}\delta^{10}(k_1+k_2+k_3)
\,\sqrt {2\alpha '}\,
\epsilon_1^\alpha\epsilon_2^\mu\epsilon_3^\lambda\, t_{\alpha\mu\lambda}(k_i) 
\,.\eea
We see that the Neveu-Schwarz computation of $A_3$ has no 
$\alpha'$ corrections, in contrast with (\ref{eq:bosglu}).
Non-renormalization theorems for the superstring often prevent
$\alpha'$ corrections to the tree level three point functions, 
although not to four point functions. 
For both expressions, the zero slope limit reduces to the
conventional three gluon field theory coupling:
$\lim_{\alpha'\rightarrow 0} {A_3}{1\over{\sqrt{2\alpha'}}}
= (2\pi)^d \delta^{10}(\sum_i k_i)
\epsilon_1^\alpha\epsilon_2^\mu\epsilon_3^\lambda\, t_{\alpha\mu\lambda}(k_i)
\,.$

The space of states for the open {\it superstring},
$\tilde{\cal H}$,
is obtained by starting with the states of the untwisted Neveu-Schwarz
theory, ${\cal H}$, introduced above, then adding in 
keeping only the subspace of each
defined by $\theta =1$, with $\theta^2=1$.
The states of the untwisted theory
are generated by the action of $d$ infinite sets of half-integrally moded
oscillators, $b^\mu_s$, $0\leq \mu\leq d-1$ 
(together with the integrally moded
oscillators, $a^\mu_n$),
on the vacuum state, $|0\rangle$.
The twisted sector is obtained from the action of $d$ infinite sets of
integrally moded oscillators, $d^\mu_n$, (together with the integrally moded
oscillators, $a^\mu_n$), 
on the twisted ground states which form
a $2^{d/2}$ irreducible representation,
$\X$, of the gamma matrix Clifford algebra, $\{\gamma^\mu\}$.
The involution $\theta$ is defined on the untwisted space ${\cal H}$ by
$\theta |0\rangle = (-1)|0\rangle$, $\theta b_s^\mu\theta^{-1} = -b_s^\mu$,
and on the twisted space, ${\cal H}_T$, by
$\theta |0\rangle_R^{\pm} = \pm|0\rangle_R^{\pm}$, $\theta d_n^\mu\theta^{-1} = -d_n^\mu$,
where $\X = |0\rangle_R^+ + |0\rangle_R^-$ 
Whenever d is even we can define
$\gamma^{d+1} \equiv \gamma^1\gamma^2\ldots\gamma^d$ which satisfies
$\{\gamma^{d+1},\gamma^\mu\} = 0$, $(\gamma^{d+1})^2 = 1$.
The operators $\half (1\pm\gamma^{d+1}(-1)^{\sum_{n>0}d_{-n}\cdot d_n})$ 
are chirality
projection operators; they project onto spinors of definite chirality.
A spinor of definite chirality is called a Weyl spinor; and the restriction
to spinors of one chirality or the other is called a Weyl condition.
In the Neveu-Schwarz sector, $\theta\equiv (-1) 
(-1)^{\sum_{s>0}b_{-s}\dt b_s}$,
and in the Ramond sector, $\theta \equiv \gamma^{d+1} (-1)^{\sum_{n>0}d_{-n}
\dt d_n}$.
The Ramond Fock space splits into two  
$SO(d)$ invariant subspaces, according to the eigenvalue of $\theta$, the
ground states being denoted by $|0\rangle_R^\pm$. The $\theta = 1$ subspace
is thus a projection onto the odd $b$ sector, and onto chiral fermions in
the Ramond sector, and is known as the {\it Gliozzi-Scherk-Olive} (GSO)
projection. 
The worldsheet fermion fields are $\psi^\mu(z)$ in either
the Neveu-Schwarz $\psi^\mu(z) = b^\mu(z)$ or 
Ramond representation $\psi^\mu(z) = d^\mu(z) = \sum_n d_n^\mu z^{-n-\half}$,
representing the vertices for emitting the massless vector state
from a Neveu-Schwarz or Ramond line, respectively. 
The ground state $|a\rangle$ of the Ramond sector of the superstring
is a spacetime fermion which is in one-to-one correspondence with
a worldsheet field $S_a(z)$ called a {\it spin field}:
\be |a\rangle = \lim_{z\rightarrow 0} S_a(z) |0\rangle\,.\ee
The spin fields are non-local with respect to the ordinary
superconformal fields $\psi^\mu(z)$:
\bea
&\psi^\mu(z) S^a (\zeta) \sim
(z-\zeta)^{-\half} \gamma^{\mu\,a}_{\hskip5pt\bdot}
S^{\bdot}(\zeta)\cr
&\psi^\mu(z) \psi^\nu(\zeta) \sim
(z-\zeta)^{-1}\cr
&a^\mu(z) a^\nu(\zeta) \sim (z-\zeta)^{-2}\,,\quad
a^\mu(z)\psi^\nu(\zeta) \sim 0\,,\quad
a^\mu(z) S^a(\zeta)\sim 0\cr
\eea
due to the non-meromorphic structure of the operator
product of $\psi$ with $S$. 
It follows that the worldsheet supercurrents $G(z)$
are not local with respect to the spin fields. It follows that 
in the presence of {\it background Ramond fields}, the
superconformal invariance of the worldsheet action $S_2$
would appear to be violated: 
the supersymmetrization of the bosonic Polyakov action (\ref{eq:wsa})
contains terms of the form
\be
S_2 = -{1\over 4\pi\alpha '} \int d\sigma d\tau  \sqrt{|g|}
(\,\partial_\alpha x^\mu\partial_\beta\,x^\nu 
+ i\bar\psi^\mu \gamma^\beta \partial_\beta\psi^\nu ) \,G_{\mu\nu}(x)
\ee
and is superconformally invariant, called the 
guiding principle of the
$RNS$ description of perturbative superstrings.  
When a background Ramond field ${\cal B}_{\mu\nu}$
is required, one
might try to add to $S_2$ a term such as
\be
\int d\sigma d\tau S_a^L S_b^R [\gamma^\mu,\gamma^\nu ]^{a b}
B_{\mu\nu}(x)\ee
but the presence of spin fields jeopardizes the 
{\it super}conformal invariance.
Several efforts to answer this problem have been 
made, some of which we cite here~\cite{tsy,ts,kt,dg,bvw,berkthree}.
One formulation~\cite{bvw,berkthree} which survives quantization
dispenses with spin fields altogether. It is 
discussed in sections 3.3 and 4.

An extension of the $RNS$ superstring to include BRST ghost fields
recasts the physical state conditions as cohomology. 
In this reformulation, in addition to the left and right moving
``matter'' fields $X^\mu(z,\bar z)$,
$\psi^\mu_L(z),\psi^\mu_R(\bar z)$ combining to give central
charge $c=15$, 
there are ``ghost'' fields
$b_L(z),c_L(z), \beta_L(z),\gamma_L(z)$ and 
$b_R(\bar z),c_R(\bar z), \beta_R(\bar z),\gamma_R(\bar z)$
contributing $c=-15$.
Left and right-moving BRST charges each have the
structure $Q \sim c (L_m + \half L_g) + \gamma (G_m + \half G_g)$
and the left and right Virasoro generators have
zero central charge $L \sim L_m + L_g $.

\subsection{Green-Schwarz ($GS$)}

This is a supersymmetric but not Lorentz covariant quantization.
The open string worldsheet fields are $S_a(z)$,
$a^\mu(z)$ and satisfy meromorphic operator products
\be S^a(z) \bar S^b(\zeta) \sim (z-\zeta)^{-1} c^{ab}\,,\quad
a^i(z) a^j(\zeta) \sim (z-\zeta)^{-2} \delta^{ij}\,,\quad
a^i(z) S^a(\zeta)\sim 0 \ee
since the coordinates are limited to the light-cone
$1\le i,j\le 8$, and $1\le a,b\le 8$.
Here the $RNS$ spin field $S^a(z)$ has been promoted to
a fundamental worldsheet variable, whereas in the 
$RNS$ case in fact it is expressible as a combination of
the $\psi$ fields, $S\sim e^{b\ldots d}$. The Green-Schwarz
formulation of the superstring dispenses with the need to sum over 
different spin stuctures (related to the NS and R sectors)
in the one-loop string amplitudes. 

\subsection{Berkovits-Vafa-Witten ($BVW$)}

This is a covariant and supersymmetric quantization in six spacetime
dimensions. It has been applied primarily to compactifications
of the Type IIB string either in the ``flat'' case
$R^6\times K3$, or the ``curved'' case $AdS_3\times S^3\times K3$.
$BVW$ provides a partially covariant quantization of the
Green-Schwarz superstring.
Eight of the sixteen supersymmetries are manifest,
in the sense that they act geometrically on the target space
of the worldsheet sigma model. In addition, there are no
worldsheet spin fields and so can more easily incorporate
Ramond-Ramond background fields. 

The $BVW$ worldsheet fields are
$X^m, \theta^a,\bar\theta^a$, the conjugate fermions
$p^a,\bar p^a$ for $1\le m\le 6; 1\le a\le 4$,
and two additional worldsheet bosons
$\rho,\sigma,\bar\rho,\bar\sigma$. These describe the
$d=6$ part of the Type IIB string. The $K3$ part is
described by the standard $RNS$ description of a
$T^4/Z_2$ orbifold.  
The $\theta^a$'s are ordinary conformal fields, not spin fields.
In flat space, the worldsheet variables are holomorphic
and satisfy free operator products relations including
\be p_a(z) \theta^b(\zeta)  \sim (z-\zeta)^{-1}\delta^b_a\,.\ee
In curved space, the worldsheet fields are no longer holomorphic,
and the worldsheet action becomes a sigma model with the supergroup
$PSU(2|2)$ as target, which is no longer a free conformal field theory
nor a Wess-Zumino-Witten ($WZW$) model. 

A ten dimensional version of these variables has appeared 
recently~{\cite{berkthree}-\cite{bli}}. 
In flat spacetime, field redefinitions give back
the RNS formalism. In $AdS_5\times S^5$, vertex operator
constraint equations have been considered~\cite{bc}. 
The Berkovits variables
are given by the ten-dimensional superspace variables
$X^\mu(z,\bar z)$, $\theta^\alpha_L (z,\bar z)$,
$\theta^\alpha_R (z,\bar z)$, for $0\le\mu\le 9$,
$1\le\alpha\le 16$; and the conjugate fermionic worldsheet fields
$p^\alpha_L (z,\bar z)$, $p^\alpha_R (z,\bar z)$.
There are additional worldsheet bosons that are
spacetime spinors $\lambda^\alpha_L(z,\bar z),
\lambda^\alpha_R(z,\bar z),$ which separately
satisfy $\lambda^\alpha \gamma^\mu_{\alpha\beta} \lambda^\beta = 0$
and carry $22$ degrees of freedom. 
The construction includes
left and right-moving BRST charge operators and
Virasoro generators. The contribution to the central charge
is $10 + 22$ from the worldsheet bosons, and $-32$ from the
worldsheet fermions. The variables do not exhibit
the ``matter times ghost'' structure of conventional the BRST 
formalism. For $AdS_5\times S^5$ spacetime, 
the worldsheet fields are not holomorphic.  

\section{Type IIB Superstrings on $AdS_3\times S^3\times K3$}

Compatification of the Type IIB superstring on
either $R^6\times K3$ or $AdS_3\times S^3\times K3$ 
yields a $d=6$, $N=(2,0)$ theory which has sixteen supercharges.
The $6d$ massless particle content is a supergravity multiplet
and 21 tensor multiplets. In flatspace the multiplets are representations
of the light-cone little group $SO(4)$:
sg $(3,3) + 5(3,1) + 4 (3,2)$ , tensor $(1,3) + 5(1,1) + 4 (1,2)$.
In curved space the number of physical
degrees of freedom in the multiplets remains the same. 
The ``compatification independent'' 6d fields
make up the supergravity multiplet and one of the tensor multiplets,
they are the graviton $(3,3)$, an antisymmetric tensor $(3,1) + (1,3)$, 
and a scalar $(1,1)$: $g_{mn}(x), b_{mn}(x), \phi(x)$;
four self-dual tensors $(3,1)$ and four scalars contained in
$V_{aa}^{--}(x), F^{++}(x), A^{-+\bar a}_a(x), A^{+-a}_{\bar a}$;
and four gravitinos $(3,2)$ and four spinors $(1,2)$ contained in
$\xi^-_{ma}(x),\bar\xi^-_{ma}(x),\chi^{a}_m(x),\bar \chi^{a}_m(x)\,.$ 

The vertex operator which describes these states is given in 
terms of the (compactification independent) worldsheet fields
$X^m,\theta^a,\bar\theta^a$ by the superfield
\begin{eqnarray}
V_{1,1}&=&\t^a\bar\t^{\bar a} V^{--}_{a\bar a} +
\t^a\t^b\bar\t^{\bar a}\s^m_{ab} \bar\xi^-_{m\,\bar a}+
\t^a\bar\t^{\bar a} \bar\t^{\bar b}\s^m_{\bar a\bar b}
\xi^-_{m\, a}\nonumber\\
&\hskip15pt +&\t^a\t^b\bar\t^{\bar a} 
\bar\t^{\bar b}\s^m_{ab}\s^n_{\bar a\bar b}
( g_{mn}+b_{mn}+\bar g_{mn}\phi) +
\t^a(\bar\t^3)_{\bar a} A^{-+\,\bar a}_{a}
+(\t^3)_a\bar\t^{\bar a} A^{+-\, a}_{\bar a}\nonumber\\
&\hskip15pt +&\t^a\t^b(\bar\t^3)_{\bar a}\s^m_{ab} \bar\chi_m^{+\,\bar a}+
(\t^3)^a\bar\t^{\bar a} \bar\t^{\bar b}
\s^m_{\bar a\bar b} \chi_m^{+\, a}+
(\t^3)_a(\bar\t^3)_{\bar a} F^{++\, a\bar a}\,.\nonumber\\
\end{eqnarray}
The string constraint equations which select the physical states
are generated in this formalism by a topological $N=4$ superVirasoro algebra
we will discuss in the next section. For flat spacetime,
the constraints will result in that all the above 6d fields satisfy 
$\partial^m\partial_m\phi=0$ and
\begin{eqnarray}
\p^m g_{mn} &=& -\p_n\phi\,,\quad  \p^m b_{mn} = 0\,,\quad
\p^m \chi^{\pm b}_m = \p^m \bar\chi^{\pm\bar b}_m = 0\nonumber\\
\partial_{ab} \chi_m^{\pm b} &=& \partial_{\bar a\bar b}
\chi_m^{\pm \bar b} = 0\,,\quad
\partial_{cb} F^{\pm\pm b\bar a} =
\partial_{\bar c\bar b} F^{\pm\pm \bar b a} = 0\,,
\label{eq:flateom}\end{eqnarray}
where
\begin{eqnarray}
F^{+- a\bar a} = \p^{\bar a\bar b}
A_{\bar b}^{+- a}&\,,&\quad
F^{-+ a\bar a} = \p^{a b}
A_b^{-+ \bar a}\,,\quad
F^{-- a\bar a} = \p^{a b} \p^{\bar a\bar b} V_{b\bar b}^{--}\cr
&\chi_m^{-a}& = \p^{ab} \xi_{m b}^-\,, \quad
\bar\chi_m^{- \bar a} = \p^{\bar a\bar b} \bar\xi_{m{\bar b}}^-\,.
\end{eqnarray}
These are equivalent to the equations of motion
for $D=6$, $N=(2,0)$ supergravity~\cite{lr} with one tensor multiplet
expanded around the six-dimensional Minkowski metric.

In the curved case $AdS_3\times S^3$, the constraints will
result in a different set of equations of motion for the 6d fields.
We give the answer here for the bosonic 6d fields, 
and show the derivation in section 4.2.
The six-dimensional metric field $g_{rs}$,
the dilaton $\phi$, and the two-form $b_{rs}$ satisfy
\begin{eqnarray}
{\textstyle{1\over 2}} D^p D_p b_{rs} &=&
-{\textstyle{1\over 2}} (\sigma_r\sigma^p\sigma^q)_{ab}\delta^{ab}
\,D_p\,
[\,g_{qs} + \bar g_{qs}\phi\,]
+{\textstyle{1\over 2}} (\sigma_s\sigma^p\sigma^q)_{ab}\delta^{ab}
D_p\,
[\,g_{qr} + \bar g_{qr}\phi\,]\nonumber\\
&-&\bar R_{\tau r s \lambda} \, b^{\tau\lambda}\,
-{\textstyle{1\over 2}} \bar R_r^{\,\,\tau}\, b_{\tau s}
-{\textstyle{1\over 2}} \bar R_s^{\,\,\tau}\, b_{r\tau}\nonumber\\
&+& {\textstyle{1\over 4}} F^{++gh}_{\rm asy}
\,\sigma_r^{ab}\sigma_s^{ef}\, \delta_{ah}\delta_{be}\delta_{gf}
\label{eq:steqone}\end{eqnarray}
\begin{eqnarray}
{\textstyle{1\over 2}} D^p D_p \,
(\, g_{rs} + \bar g_{rs} \phi \,)\,&=&
-{\textstyle{1\over 2}} (\sigma_r\sigma^p\sigma^q)_{ab}\delta^{ab}
\,D_p b_{qs} \,
+ {\textstyle{1\over 2}} (\sigma_s\sigma^p\sigma^q)_{ab}\delta^{ab}
\,D_p b_{rq} \nonumber\\
&-& \bar R_{\tau r s \lambda} \,
(\, g^{\tau\lambda} + \bar g^{\tau\lambda}\phi \,)\,
-{\textstyle{1\over 2}} \bar R_r^{\,\,\tau}\,
(\, g_{\tau s} + \bar g_{\tau s} \phi \,)\nonumber\\
&-& 
{\textstyle{1\over 2}} \bar R_s^{\,\,\tau}\,
(\, g_{r\tau } + \bar g_{r\tau }\phi \,) +
{\textstyle{1\over 4}} F^{++gh}_{\rm sym} \,\sigma_{rga}\sigma_{shb}\,
\delta^{ab}\,.\nonumber\\
\label{eq:steqtwo}\end{eqnarray}
This is the curved space version of the flat space zero Laplacian condition
$\p^p \p_p b_{rs} = \p^p \p_p g_{rs} = \p^p \p_p \phi = 0$.

Four self-dual tensor and scalar pairs come from the string bispinor
fields\break
$ F^{++ ab}, \,V^{--}_{ab}, \,,A^{+- a}_b, \,A^{-+ b}_a$.
From the string constraint equations they satisfy
\begin{equation}
\sigma^p_{da}\, D_p \, F^{++ab}_{\rm asy} = 0
\end{equation}
\begin{equation} 
{\textstyle{1\over 4}}\,[\, \delta^{Ba} \,\sigma^r_{ga}\,D_r\,
F^{++gH}_{\rm sym} \, - \,
\delta^{Ha} \,\sigma^r_{ga}\,D_r\, F^{++gB}_{\rm sym} \, ]
= -{\textstyle{1\over 4}}\,\, \epsilon^{BH}_{\qquad cd}\,\,
F^{++cd}_{\rm asy}
\end{equation}
These can be shown~\cite{dw} to be 
equivalent to the linearized supergravity equations~\cite{sez}
for the supergravity multiplet and one tensor multiplet
of the $d=6$, $N=(2,0)$ theory expanded around the
$AdS_3\times S^3$ metric and a self-dual three-form,
by using the following field identifications:
the  vertex operator components 
in terms of the supergravity fields
$g_{prs}, g^6_{prs}, h_{rs}, \phi^i$ , $1\le i\le 5$,
(and $2\le I\le 5$ ) are
\be
H_{prs}\equiv  \,g^6_{prs} + 2 \,g^1_{prs}\,+ B^I \,g^I_{prs}\ee
\be g_{rs} \equiv \,h_{rs}
-{\textstyle{1\over 6}}\bar g_{rs}\, h^\lambda_{\,\,\lambda}\ee
\be\phi = -{\textstyle{1\over 3}}\, h^\lambda_{\,\,\lambda}\ee
\bea
&F^{++ab}_{\rm sym} =  \,{\textstyle{2\over 3}}
(\sigma_p\sigma_r\sigma_s)^{ab}
\, B^I \,\,g^I_{prs} +  \delta^{ab}\,\phi^{++}\cr
&F^{++ab}_{\rm asy} = \,\,\sigma^{p\,ab} \, D_p\,\phi^{++}\cr
&\phi^{++} =  4 C^I\,\,\phi^I
\eea
which follows from choosing the graviton trace
$h^\lambda_{\hskip3pt\lambda}$ to satisfy $\phi^1 -
h^\lambda_{\,\,\lambda} \,\equiv\, - 2\, C^I\phi^I$.
Here $H_{prs} \equiv \partial_p b_{rs} +  \partial_r b_{sp} +
\partial_s b_{pr}$.
The combinations $C^I\phi^I$ and $B^I g^I_{prs}$ reflect the
$SO(4)_{\rm R}$ symmetry of the $D=6, N=(2,0)$ theory on $AdS_3\times S^3$.
We relabel $C^I = C^I_{++}$, $B^I = B^I_{++}$.
To define the remaining string components in terms of supergravity fields,
we consider linearly independent quantities $C_\ell^I\phi^I$,
$B_\ell^I  g^I_{prs}$, $\ell = ++,+-, -+, --$.

\bea
&F^{+-ab}_{\rm sym} =  \,\, {\textstyle{2\over 3}}
(\sigma_p\sigma_r\sigma_s)^{ab}
\, B_{+-}^I \,\,g^I_{prs} +  \delta^{ab}\,\phi^{+-}\cr
&F^{+-ab}_{\rm asy} = \,\,\sigma^{p\,ab} \, D_p\,\phi^{+-}\cr
&\phi^{+-} =  4 C_{+-}^I\,\,\phi^I\cr
&F^{-+ab}_{\rm sym} =  \,\, {\textstyle{2\over 3}}
(\sigma_p\sigma_r\sigma_s)^{ab}
\, B_{-+}^I \,\,g^I_{prs} +  \delta^{ab}\,\phi^{-+}\cr
&F^{-+ab}_{\rm asy} = \,\,\sigma^{p\,ab} \, D_p\,\phi^{-+}\cr
&\phi^{-+} = 4 C_{-+}^I\,\,\phi^I
\eea
$V^{--}_{ab}$ is given in terms of the fourth tensor/scalar pair
$C_{--}^I\,\,\phi^I$, $B_{--}^I g_{mnp}^I$ through
\be D^p D_p\, V_{cd}^{--}
\, - \delta^{gh}\sigma^p_{ch}\, D_p \, V_{gd}^{--}
\, + \delta^{gh}\sigma^p_{dh}\, D_p \, V_{cg}^{--}
\, +{\textstyle{1\over 2}} \epsilon_{cd}^{\,\,\,\,gh}\, V_{gh}^{--}
=\,-8\,\sigma^m_{ce}\,\sigma^n_{df}\,\delta^{ef}\, G_{mn}\,.\ee

\subsection{Topological Strings}

The origin of the constraints is an $N=4$ twisted superconformal algebra.
In this section, we review how the superstring can be reformulated 
as an $N=4$ topological string theory, and show how this formalism gives a 
description of the superstring with manifest $d=6$ spacetime supersymmetry.
That is to say, the spectrum of the superstring can be identified
with the states surviving a set of $N=4$ constraints. 
We begin this subsection by remembering how the bosonic string
can be reorganized as an $N=2$ topological string~\cite{bv,bvw}.
An $N=2$ topological string has a twisted $N=2$ superconformal algebra 
\bea
&\tilde T(z) \tilde T(\zeta) = 
(z-\zeta)^{-2} 2 \tilde T(\zeta) + (z-\zeta)^{-1} \partial \tilde T(\zeta)\,,
\cr
&\tilde T(z) G^+ (\zeta) = (z-\zeta)^{-2} G^+ (\zeta)
+ (z-\zeta)^{-1} \partial G^+(\zeta)\,,\cr
&\tilde T(z) G^- (\zeta) = (z-\zeta)^{-2} 2 G^- (\zeta)
+ (z-\zeta)^{-1} \partial G^-(\zeta)\,,\cr
&G^+(z) G^-(\zeta) = (z-\zeta)^{-3} {c\over 3} 
+ (z-\zeta)^{-2} J(\zeta) 
+ (z-\zeta)^{-1}  \tilde T(\zeta) \,,\cr
&G^+(z) G^+(\zeta) = 0\,, &*\cr
&G^-(z) G^-(\zeta) = 0\,,\cr
&\tilde T(z) J(\zeta) = (z-\zeta)^{-3} (-{c\over 3}) 
+ (z-\zeta)^{-2} J(\zeta) 
+ (z-\zeta)^{-1} \partial J(\zeta)\,,\cr
&J(z) J(\zeta) = (z-\zeta)^{-2} {c\over 3}\,,&*\cr
&J(z) G^+(\zeta) = (z-\zeta)^{-1} G^+(\zeta)\,,&*\cr
&J(z) G^-(\zeta) = -(z-\zeta)^{-1} G^-(\zeta)\,.
\label{eq:n2tw}\eea
(\ref{eq:n2tw})
is related to the generators of the (untwisted)
$N=2$ superconformal algebra :
\bea
&L(z) L(\zeta) = (z-\zeta)^{-4} {c\over 2} 
+ (z-\zeta)^{-2} 2 L(\zeta) + (z-\zeta)^{-1} \partial L(\zeta)\,,\cr
&L(z) G^\pm (\zeta) = (z-\zeta)^{-2} \thalf G^\pm (\zeta)
+ (z-\zeta)^{-1} \partial G^\pm(\zeta)\,,\cr
&G^+(z) G^-(\zeta) = (z-\zeta)^{-3} {c\over 3} 
+ (z-\zeta)^{-2} J(\zeta) 
+ (z-\zeta)^{-1} ( L(\zeta) +\half\partial J(\zeta) )\,,\cr
&G^+(z) G^+(\zeta) = 0\,,\quad
G^-(z) G^-(\zeta) = 0\,,\cr
&L(z) J(\zeta) = (z-\zeta)^{-2} J(\zeta) 
+ (z-\zeta)^{-1} \partial J(\zeta)\,,\cr
&J(z) J(\zeta) = (z-\zeta)^{-2} {c\over 3}\,,\cr
&J(z) G^\pm(\zeta) = \pm (z-\zeta)^{-1} G^\pm(\zeta)\,,\eea
where the generators differ only by the twisted
Virasoro generator
$\tilde T(z) \equiv L(z)  +{\textstyle{1\over 2}} \partial J(z)$.
Since the OPE's (\ref{eq:n2tw}) resemble somewhat those of the bosonic string
(\ref{eq:n2bs})
we can define physical fields relative to $Q_0$ cohomology
where $G^+(z) = \sum_n Q_n z^{-n-1}$. Physical fields
correspond to chiral primary fields $\Phi^+(z)$ with ghost charge $+1$ and
dimension $0$ (they arise from operators that have ghost charge $+1$ and
dimension $\half$ before the algebra is twisted), so that
$\{Q_0, \Phi^+(\zeta)\} = 0$.

Bosonic string theory can be viewed as 
a two-dimensional conformal
field theory with certain additional features (we concentrate on
left-movers and will denote right-movers with barred notation):
\bea
&T_{\rm tot}(z) T_{\rm tot}(\zeta) = 
(z-\zeta)^{-2} \,2 T_{\rm tot} (\zeta) + (z-\zeta)^{-1} 
\partial T_{\rm tot}(\zeta)\,,\cr
&T_{\rm tot} (z) j_{\rm BRST}(\zeta) = (z-\zeta)^{-2}  j_{\rm BRST}(\zeta)
+ (z-\zeta)^{-1} \partial j_{\rm BRST}(\zeta)\,,\cr
&T_{\rm tot} (z) b(\zeta) = (z-\zeta)^{-2}  2 b(\zeta)
+ (z-\zeta)^{-1} \partial b(\zeta)\,,\cr
&j_{\rm BRST}(z) b(\zeta) = (z-\zeta)^{-3} {(c=9)\over 3} 
+ (z-\zeta)^{-2} j_{\rm ghost}(\zeta) 
+ (z-\zeta)^{-1}  T_{\rm tot}(\zeta) \,,\cr
&j_{\rm BRST}(z) j_{\rm BRST}(\zeta) = -  j_{\rm BRST}(\zeta) j_{\rm BRST}(z)
= 2{\partial\over {\partial\zeta}} 
[ (z-\zeta)^{-2} \nox \partial c(\zeta) \, c(\zeta) \nox] \ne 0\, \cr 
&b(z) b(\zeta) = 0\,,\cr
&T_{\rm tot} (z) j_{\rm ghost} (\zeta) = (z-\zeta)^{-3} {(c=9)\over (-3)}
+ (z-\zeta)^{-2} j_{\rm ghost}(\zeta) 
+ (z-\zeta)^{-1} \partial j_{\rm ghost}(\zeta)\,,\cr
&j_{\rm ghost}(z) j_{\rm ghost}(\zeta) = (z-\zeta)^{-2} {(c=3)\over 3}\,,\cr
&j_{\rm ghost}(z) j_{\rm BRST}(\zeta) 
=  (z-\zeta)^{-3} 4 c(\zeta) +  (z-\zeta)^{-2} 2\partial c(\zeta) 
+ (z-\zeta)^{-1} j_{\rm BRST}(\zeta)\,,\cr
&j_{\rm ghost}(z) b(\zeta) = -(z-\zeta)^{-1} b(\zeta)\,,
\label{eq:n2bs}\eea
where the generators are
\bea
&T_{\rm tot} = T_m^{N=0} + T_g^{N=0}
= T_m^{N=0} -2\nox b\partial c\nox - \nox\partial b\, c\nox\cr
&j_{\rm BRST} (z) = 
c T_m +\hhalf \nox c T_g\nox + {\thalf}\partial^2 c
= c T_m - \nox c b \partial c\nox + {\thalf}\partial^2 c\cr
&b(z) \cr
&j_{\rm ghost} (z) = \nox cb \nox 
\eea
and $c(z) b(\zeta) = - b(\zeta) c(z) =  (z-\zeta)^{-1} 
+ \nox c(z) b(\zeta)\nox$. The normal ordering has been defined
putting the annihilation operators to the right of the creation
operators, where $b_n |0\rangle^{bc} = 0$ for $n\ge -1$;
$c_n |0\rangle^{bc} = 0$ for $n\ge 2$. The BRST charge is
$Q\equiv {1\over{2\pi i}}\oint dz j_{BRST}(z)
= \sum_m c_m L_m^X -\hhalf\sum_{m,n} (m-n) 
\nox c_{-m} c_{-n} b_{n+m}\nox $. Then $Q \, (|0\rangle^{bc}\otimes
|0\rangle_X ) \,=0$. The ghost charge is
$J_0\equiv \sum_n\nox c_n b_{-n}\nox $. Then
$(c_1 |0\rangle^{bc}\otimes |\phi\rangle_X )$ has ghost charge eigenvalue,
{\it i.e.} ghost number, equal to one.
The physical state conditions in the ``old covariant'' formalism
are $(L^X_0 - 1) |\phi\rangle_X = 0 $, $L^X_n |\phi\rangle_X = 0$ for $n>0$.
Since $Q |\psi\rangle = 0$ implies $ (c_0 (L^X_0 -1) + \sum_n c_{-n} L^X_n )
\,|\psi\rangle = 0$
when $|\psi\rangle = c_1 |0\rangle^{bc}\otimes |\phi\rangle_X$,
then in the BRST formalism
the physical vertex operators are defined by ghost number one fields
$\Phi^+(z)$ that obey $\{Q, \Phi^+ (z)\} = 0$, {\it i.e.} the OPE of
$j_{BRST}(z) \Phi^+(\zeta)$ has no single pole.
So here, every physical state is in one-to-one correspondence with
a primary
field of the Virasoro algebra of dimension 0, {\it i.e.}
$\Phi^+(z) = c(z) \phi_X(z)$.
{\it Thus we have used the twisted $N=2$ superVirasoro algebra to define
physical fields relative to $Q_0$ cohomology where
$G^+(z) = \sum_n Q_n z^{-n-1}$. Physical states 
correspond to chiral primary fields 
$\Phi^+(z)$ with ghost charge $+1$ and dimension$0$}.

{}For $N=2$ topological strings, $c=9$. $N=4$ 
topological strings are used when $c=6$.
{}From an $N=2$ superconformal algebra with $c=6$, we construct
a topological $N=4$ string by defining the remaining generators
and twisting the $N=4$ superconformal algebra:
\bea
&\tilde T(z) \tilde T(\zeta) = 
(z-\zeta)^{-2} 2 \tilde  T(\zeta) + (z-\zeta)^{-1} 
\partial \tilde T (\zeta)\,,\cr
&\tilde T(z) G^+ (\zeta) = (z-\zeta)^{-2} G^+ (\zeta)
+ (z-\zeta)^{-1} \partial G^+(\zeta)\,,\cr
&\tilde T(z) G^- (\zeta) = (z-\zeta)^{-2} 2 G^- (\zeta)
+ (z-\zeta)^{-1} \partial G^-(\zeta)\,,\cr
&G^+(z) G^-(\zeta) = (z-\zeta)^{-3} {(c=6)\over 3} 
+ (z-\zeta)^{-2} J(\zeta) 
+ (z-\zeta)^{-1}  \tilde T(\zeta) \,,\cr
&G^+(z) G^+(\zeta) = 0\,, \quad\qquad
G^-(z) G^-(\zeta) = 0\,,\cr
&\tilde T (z) J(\zeta) = (z-\zeta)^{-3} (-{(c=6)\over 3}) 
+ (z-\zeta)^{-2} J(\zeta) 
+ (z-\zeta)^{-1} \partial J(\zeta)\,,\cr
&J(z) J(\zeta) = (z-\zeta)^{-2} {(c=6)\over 3}\,,\cr
&J(z) G^\pm(\zeta) = \pm (z-\zeta)^{-1} G^\pm(\zeta)\,,\cr
&J(z) J^\pm(\zeta) = (z-\zeta)^{-1} (\pm 2) J^\pm(\zeta)\,,\cr
&J^+(z) J^-(\zeta) = - (z-\zeta)^{-2} {(c=6)\over 6} 
- (z-\zeta)^{-1} J(\zeta)\,,\cr
&\tilde T(z) J^+ (\zeta) = 
(z-\zeta)^{-1} \partial J^+(\zeta)\,,\cr
&\tilde T(z) J^- (\zeta) = 
(z-\zeta)^{-2} 2 J^-(\zeta) 
+ (z-\zeta)^{-1} \partial J^-(\zeta)\,,\cr
&J^+(z) G^-(\zeta) = - (z-\zeta)^{-1} \tilde G^+(\zeta)\,,\qquad
J^-(z) G^-(\zeta) = 0\,,\cr
&J^-(z) G^+(\zeta) = - (z-\zeta)^{-1} \tilde G^-(\zeta)\,,\qquad
J^+(z) G^+(\zeta) = 0\,,\cr
&J^-(z) \tilde G^+(\zeta) = (z-\zeta)^{-1} G^-(\zeta)\,,\qquad
J^-(z) \tilde G^-(\zeta) = 0\,,\cr
&J^+(z) \tilde G^-(\zeta) = (z-\zeta)^{-1} G^+(\zeta)\,,\qquad
J^+(z) \tilde G^+(\zeta) = 0\,,\cr
&G^-(z) \tilde G^+(\zeta) =0\,,\qquad
G^+(z) \tilde G^-(\zeta) =0\,,\cr
&\tilde G^+(z) \tilde G^-(\zeta) = (z-\zeta)^{-3} {(c=6)\over 3} 
+ (z-\zeta)^{-2} J(\zeta) 
+ (z-\zeta)^{-1}  \tilde T(\zeta) \,,\cr
&G^+(z) \tilde G^+(\zeta) = (z-\zeta)^{-2} 2 J^+(\zeta) 
+ (z-\zeta)^{-1} \partial J^+(\zeta)\,,\cr
&G^-(z) \tilde G^-(\zeta) = (z-\zeta)^{-2} 2 J^-(\zeta) 
+ (z-\zeta)^{-1} \partial J^-(\zeta)\,,\cr
&\tilde G^+(z) \tilde G^+(\zeta) =0\,,\quad
\tilde G^-(z) \tilde G^-(\zeta) =0\cr
&\tilde T(z) \tilde G^+(\zeta) = (z-\zeta)^{-2} \tilde G^+(\zeta)
+ (z-\zeta)^{-1} \partial \tilde G^+(\zeta)\,,\cr
&\tilde T(z) \tilde G^-(\zeta) = (z-\zeta)^{-2} 2\tilde G^-(\zeta)
+ (z-\zeta)^{-1} \partial \tilde G^-(\zeta)\,,\cr
&J(z) \tilde G^\pm(\zeta) = \pm (z-\zeta)^{-1} \tilde G^\pm(\zeta)\,.
\label{eq:n4tw}\eea
Since the superstring can be written as an
$N=2$ super Virasoro algebra with $c=6$, it's necessary to
find additional generators making up an $N=4$ topological string.
In $RNS$ variables they are
\bea
&\tilde T(z) = T_m^{N=1} + T_g^{N=1} 
= T_m^{N=1} - 2\nox b\partial c\nox -\nox \partial b c\nox 
-\thalf\nox\beta\partial \gamma\nox -\half\nox\partial\beta\gamma\nox\cr
&G^+(z)= \gamma G_m + c( T_m -{\textstyle{3\over 2}}\beta\partial\gamma
-{\textstyle{1\over 2}}\partial\beta\gamma - b\partial c ) -\gamma^2 b
+\partial^2 c + \partial (c\xi\eta)\cr
&G^-(z)= b\cr
&J(z)= cb + \eta\xi\cr
&J^+(z)= c\eta\cr
&J^-(z)= b\xi\cr
&\tilde G^+(z)= \eta\cr
&\tilde G^-(z)=  b ( i e^\phi G_m +\eta e^{2\phi} \partial b - c\partial\xi)
+\xi (T_m -{\textstyle{3\over 2}}\beta\partial\gamma
-{\textstyle{1\over 2}}\gamma\partial\beta - 2 b\partial c + c\partial b )
+\partial^2\xi\,,\cr
&\label{eq:n4gen}
\eea
with $c(z) b(\zeta) = - b(\zeta) c(z) =  (z-\zeta)^{-1} 
+ \nox c(z) b(\zeta)\nox$. Also the super-reparametrization ghosts with
$\gamma(z) \beta(\zeta) =  \beta(\zeta) \gamma(z) =  (z-\zeta)^{-1} 
+ \nox \gamma(z) \beta(\zeta)\nox$ have been bosonized
as ($\beta = ie^{-\phi}\partial\xi, \gamma= -i\eta e^\phi)$
with $\xi(z) \eta(\zeta) = - \eta(\zeta) \xi(z) = (z-\zeta)^{-1} 
+\nox \xi(z) \eta(\zeta)\nox,$ and
$\phi(z) \phi(\zeta) = -\ln (z-\zeta) +\nox \phi(z) \phi(\zeta)\nox $
so that
$e^{-\phi(z)} e^{\phi(\zeta)} = :e^{-\phi(z) + \phi(\zeta)}: (z-\zeta)$.

The generators (\ref{eq:n4gen})
satisfy the $(c=6)$ twisted $N=4$ superconformal
algebra given in (\ref{eq:n4tw}). Since the algebra is twisted, the
Virasoro generators close with no anomaly ({\it i.e.} $c=0$) but
$c$ still appears in the rest of the algebra,
such as the anomaly of the $U(1)$ current $J$.
For the IIB superstring we have both the holomorphic $N=4$ superconformal
algebra (\ref{eq:n4tw}) and another anti-holomorphic one. The holomorphic 
generators, when specialized for IIB compactified to 6d,
and rewritten in terms of $BVW$ worldsheet variables which display
manifest 6d spacetime supersymmetry and eschew spin fields,
become 
\begin{eqnarray}
T&=&-\half\partial x^m\partial x_m - p_a \partial\theta^a
-\half\partial\rho\partial\rho - \half\partial\sigma\partial\sigma
+ \partial^2 (\rho + i\sigma) + T_C\nonumber\\
G^+&=&-e^{-2\rho-i\sigma} (p)^4 + {\textstyle{i\over 2}}e^{-\rho} p_a p_b
\partial x^{ab}\nonumber\\ 
&\hskip 5pt & 
+e^{i\sigma} (-\half\partial x^m\partial x_m - p_a \partial\theta^a 
-\half\partial (\rho + i\sigma) \partial (\rho + i\sigma) 
+\half\partial^2  (\rho + i\sigma) ) \,+ G^+_C\nonumber\\
G^-&=&e^{-i\sigma} + G_C^- \nonumber\\
J&=&\partial (\rho + i\sigma) + J_C\nonumber\\
\tilde G^+&=& e^{iH_C} (-e^{-3\rho-2i\sigma} (p)^4 + {\textstyle{i\over 2}}
e^{-2\rho-i\sigma} p_a p_b \partial x^{ab}\nonumber\\
&\hskip 7pt &
+ e^{-\rho} (-\half\partial x^m\partial x_m - p_a \partial\theta^a 
-\half\partial (\rho + i\sigma) \partial (\rho + i\sigma)\nonumber\\
&\hskip 7pt &
+\half\partial^2  (\rho + i\sigma) ) \,+ e^{-\rho-i\sigma} \tilde G_C^-
\nonumber\\
J^+&=& e^{\rho + i\sigma} J_C^+\nonumber\\
J^-&=& e^{-\rho - i\sigma} J_C^-\,.\nonumber\\
\label{eq:n4genbvw}\end{eqnarray}
These currents are given in terms of the left-moving bosons
$\partial x^m, \rho, \sigma$, and the left-moving fermionic worldsheet
fields $p^a,\t^a$,
where $1\le m\le 6, 1\le a\le 4$.
The conformal weights of $p_a,\t^a$ are $1$ and $0$, respectively.
We define $p^4\equiv {\scriptstyle{1\over 24}}
\epsilon^{abcd}  p_a p_b p_c p_d = p_1p_2 p_3 p_4;$
and $\partial x^{ab} = \partial x^m\,\sigma_m^{ab}$ where
$\sigma_m^{ab} \,\sigma_{nac} \,+ \sigma_n^{ab} \,\sigma_{mac} 
= \eta_{mn}\,\delta^b_c $. Here lowered indices mean
$\sigma_{mab}\equiv \half\epsilon_{abcd} \sigma_m^{cd}.$
Note that $e^\rho$ and $e^{i\sigma}$ are worldsheet fermions.
Also $e^{\rho + i\sigma}\equiv e^\rho e^{i\sigma} = - e^{i\sigma} e^\rho$.
Here $J_C\equiv i\partial H_C, J_C^+ \equiv - e^{i H_C},
J_C^-\equiv e^{-i H_C}$.
Both $\tilde T, G^\pm, J, J^\pm, \tilde G^\pm$
and the generators describing the compactification
$\tilde T_C, G_C^\pm, J_C, J_C^\pm, \tilde G_C^\pm$ satisfy the
twisted $N=4$, $c=6$,  superconformal algebra (\ref{eq:n4tw}), 
{\it i.e.} both
$\tilde T$ and $\tilde T_C$ have $c=0$. However, as seen in (\ref{eq:n4tw}) 
and (\ref{eq:n2tw}),
$c$ still appears in the twisted $N=4$ and $N=2$ algebras;
and the N=2 generators in (\ref{eq:n4genbvw}) $\tilde T, G^\pm, J$
decompose into a $c=0$
six-dimensional part and a $c=6$ compactification-dependent piece.
(That is to say, the uncompactified piece of the twisted
$N=2$ generators in (\ref{eq:n4genbvw}) satisfies
(\ref{eq:n2tw}) with $c=0$, not just for the Virasoro generator 
but also whereever
$c$ appears in (\ref{eq:n2tw}).

The other non-vanishing OPE's are
$x^m(z,\bar z) x^n(\zeta,\bar\zeta) = -\eta^{mn}\ln |z-\zeta|$;
for the left-moving worldsheet fermion fields
$\,p_a (z) \theta^b (\zeta) = (z-\zeta)^{-1}\delta_a^b$; and for the
left-moving worldsheet bosons
$\rho (z) \rho(\zeta) = - \ln (z-\zeta)\,;\,\,$
$\sigma (z) \sigma (\zeta) = -\ln (z-\zeta)\,.$
Right-movers are denoted by barred notation and have similar OPE's.

Both holomorphic and anti-holomorphic sets of generators
are used to implement the physical state conditions on the vertex
operators, a procedure~\cite{bv,berkone,bvw}
which results in a set of string constraint equations
for flat spacetime.
The notation $O_n \Phi$ denotes the pole of order $d+n$ in the OPE 
of $O$ with $\Phi$, when $O$ is a dimension d operator.
For the generators 
(\ref{eq:n4genbvw})
since $G^+$ and $\tilde G^+$ are dimension one, and nilpotent,
in analogy with the bosonic string,
the physical $N=4$ topological vertex operators $\Phi^+(z)$ are
defined by the conditions:
\be G_0^+ \Phi^+ = 0\,;\qquad \tilde G_0^+ \Phi^+ = 0\,;
\qquad ( J_0 - 1 ) \Phi^+ = 0\,.\ee
These are the physical conditions for a BRST-invariant vertex operator
in the standard RNS formalism for the superstring.
Since $\tilde G^+_0 = \eta_0$, the $\tilde G^+_0$ cohomology is
trivial, {\it i.e.} $\tilde G^+_0 \Phi^+ = 0$ implies
$\Phi^+ = \tilde G^+_0 V$. So it is always possible to define
a V satisfying
\be\Phi^+ = \tilde G^+_0 V\,; \qquad 
G^+_0 \tilde G^+_0 V = J_0 V = 0\,.\label{eq:oneel}\ee
Note that $\Phi^{++}(z)$ is a worldsheet fermion and
$V(z)$ is a worldsheet boson.
To describe the massless compactification
independent states we introduce the $U(1)$-neutral vertex operator
$V(z)$, but it is straightforward to go from $V(z)$ to the
$U(1)$ charge equal to one vertex operator $\Phi^(z)$, using the
relationship described above.
In addition to (\ref{eq:oneel}), one can use the gauge invariance
$V\sim V + G_0^+\Lambda + \tilde G_0^+\tilde\Lambda$ to further require
\be G_0^- V = \tilde G_0^- V = T_0 V = 0\,.\ee
In this gauge the physical $U(1)$-charged vertex operators $\Phi^{++}(z)$
of the {\it closed} $N=4$ toplogical string must satisfy
\bea & G_0^+ \Phi^{++} = \tilde G_0^+ \Phi^{++} =
\bar G_0^+ \Phi^{++} = \bar{\tilde G_0^+} \Phi^{++} = 0\,,\cr
&G_0^- \Phi^{++} = \tilde G_0^- \Phi^{++} = T_0 \Phi^{++} = 
(J_0 -1) \,\Phi^{++} = 0\,,\cr
&\bar G_0^- \Phi^{++} = \bar{\tilde G_0}^- \Phi^{++} = \bar T_0 \Phi^{++}
= (\bar J_0 -1)\, \Phi^{++} =0 \,. \eea
Similarly the $U(1)$-neutral vertex operator $V$ defined by
$\Phi^{++} = \tilde G_0^+ \,\bar{\tilde G}_0^+ V$ must satisfy the
conditions
\bea & G_0^+ \tilde G_0^+ V 
= \bar G_0^+ \bar {\tilde G_0^+} V = 0\,,\cr
&G_0^- V = \tilde G_0^- V = \bar G_0^- V = \bar {\tilde G_0}^- V
= T_0 V = \bar T_0 V = J_0 V = \bar J_0 V = 0\,.
\label{eq:flcon}\eea
The integrated form of the {\it closed} superstring vertex operator
$\Phi^{++}(z)$ is $\int d^2z G_{-1}^- \Phi^{++}$.
In terms of V it will be defined as
\be U = \int d^2z G_{-1}^- \bar G_{-1}^- G_0^+ \bar G_0^+ V\,.\ee

The $N=4$ topological prescription~\cite{bvw} for calculating superstring
tree-level amplitudes is
\be <\,V_1(z_1)\,(\tilde G^+_0 V_2(z_2))
(G^+_0 V_3(z_3)) \prod_{r=4}^n\int dz_r G^-_{-1} G^+_0 V(z_r)\,>\,.
\label{eq:tlamp}\ee
Note that since $V$ are $U(1)$-neutral, the amplitude (\ref{eq:tlamp}) has
operators with a total $U(1)$ charge equal to 2. This is related
to the RNS requirement that non-vanishing tree scattering amplitudes
must have total superconformal ghost charge $-2$ and total conformal
ghost charge $3$. The {\it closed} $N=4$ topological tree-level amplitudes
are given by
\be <\,V_1(z_1,\bar z_1)\,
(\tilde G^+_0 \bar {\tilde G^+_0 } V_2(z_2, \bar z_2))
(G^+_0 \bar G^+_0 V_3(z_3, \bar z_3)) 
\prod_{r=4}^n\int d^2 z_r G^-_{-1} \bar G^-_{-1} G^+_0 \bar G^+_0 
V(z_r, \bar z_r)\,>\,.\ee

\subsection{String Constraint Equations}

Using (\ref{eq:flcon}) on the general massless vertex operator
\begin{equation}
V = \sum_{m,n = -\infty}^\infty\,
e^{m(i\s + \rho) + n(i\bar\s + \bar\rho)}\,
V_{m,n} (x, \t, \bar\t)\,,\end{equation}
we find in flat spacetime the constraints from the left and right-moving
worldsheet super Virasoro algebras to be 
\begin{eqnarray}
(\N)^4 V_{1,n} &=& \N_a \,\N_b
\p^{a b} V_{1,n} = 0\nonumber\\
{\textstyle{1\over 6}}\e^{a b c d} \,\N_b \,\N_c \,\N_d
V_{1,n} &=& - i \N_b\, \p^{a b} V_{0,n}\nonumber\\
\N_a\,\N_b\, V_{0,n} - {\textstyle{i\over 2}} \e_{a b c d} \, \p^{cd}\,
V_{-1,n} &=& 0\,,\qquad \N_a \, V_{-1,n} = 0\,;\label{eq:sone}\\
\bar\N^4 V_{n,1} &=& \bar\N_{\bar a}\bar\N_{\bar b}
\bar\p^{\bar a\bar b} V_{n,1} = 0\nonumber\\
{\textstyle{1\over 6}} \e^{\bar a\bar b\bar c\bar d}\bar
\N_{\bar b}\bar \N_{\bar c} \bar \N_{\bar d}
V_{n,1} &=& -i \bar \N_{\bar b} \bar\p^{\bar a\bar b} V_{n,0}\nonumber\\
\bar\N_{\bar a}\bar \N_{\bar b} V_{n,0}
- {\textstyle{i\over 2}} \bar\e_{\bar a \bar b \bar c \bar d}
\, \bar\p^{\bar c\bar d}\, V_{n,-1} &=& 0\,,
\qquad \bar\N_{\bar a} \, V_{n,-1} = 0\label{eq:stwo}
\end{eqnarray}
\begin{equation}\p^p\p_p V_{m,n} = 0\label{eq:sthree}
\end{equation}
for $-1\le m,n\le 1$, with the notation
$\N_a = d/ d\t^a$, $\bar\N_{\bar a} = d/ d\bar\t^{\bar a}$,
$\partial^{ab} = -\sigma^{p ab}\,\p_p$.
These conditions further imply
$V_{m,n} = 0$ for $m>1$ or $n>1$ or $m<1$ or $n<1$, leaving
nine non-zero components. In fact, the independent degrees of
freedom can be shown to reside in $V_{11}$, and the 
surviving constraints yield (\ref{eq:flateom}).

In $AdS_3\times S^3$ space, we generalize~\cite{dw} the flat space 
string constraint equations (\ref{eq:sone}-\ref{eq:sthree})
as follows:
\begin{eqnarray}
F^4 V_{1,n} &=& F_a \,F_b
K^{a b} V_{1,n} = 0\nonumber\\
{\textstyle{1\over 6}}\e^{a b c d} \,F_b \,F_c \,F_d
V_{1,n} &=& - i F_b\, K^{a b} V_{0,n} +  2i F^a  V_{0,n} - E^a V_{-1,n}
\nonumber\\
F_a\,F_b\, V_{0,n} - {\textstyle{i\over 2}} \e_{a b c d} \, K^{cd}\,
V_{-1,n} &=& 0\,,\qquad F_a \, V_{-1,n} = 0\,;
\label{eq:tone}\end{eqnarray}
\begin{eqnarray}
\bar F^4 V_{n,1} &=& \bar F_{\bar a}\bar F_{\bar b}
\bar K^{\bar a\bar b} V_{n,1} = 0\nonumber\\
{\textstyle{1\over 6}} \e^{\bar a\bar b\bar c\bar d}\bar
F_{\bar b}\bar F_{\bar c} \bar F_{\bar d}
V_{n,1} &=& -i \bar F_{\bar b} \bar K^{\bar a\bar b} V_{n,0}
+ 2i \bar F^{\bar a}  V_{n,0} - \bar E^{\bar a} V_{n,-1}\nonumber\\
\bar F_{\bar a}\bar F_{\bar b} V_{n,0}
- {\textstyle{i\over 2}} \bar\e_{\bar a \bar b \bar c \bar d}
\, \bar K^{\bar c\bar d}\, V_{n,-1} &=& 0\,,
\qquad \bar F_{\bar a} \, V_{n,-1} = 0.\nonumber\\
\label{eq:ttwo}\end{eqnarray}
There is also a spin zero condition constructed from the Laplacian
\begin{equation}
(\,F_a \, E_a \,+ {\textstyle{1\over 8}}\e_{a b c d} \, K^{ab}\, K^{cd}\,)
\, V_{n,m} =
(\,\bar F_{\bar a} \, \bar E_{\bar a} \,
+ {\textstyle{1\over 8}}\bar\e_{\bar a \bar b \bar c \bar d} \,
\bar K^{\bar a \bar b}\, K^{\bar c\bar d}\,)
\, V_{n,m} = 0\,.
\label{eq:tthree}\end{equation}
We derived 
the curved space equations (\ref{eq:tone}-\ref{eq:tthree}) 
by deforming the flat space equations
by requiring invariance under the $PSU(2|2)$
transformations (\ref{eq:symce}) 
that replace the $d=6$ super Poincare transformations
of flat space. The Lie algebra of the supergroup $PSU(2|2)$
contains six even elements $K_{ab}\in SO(4)$ and eight odd
$E_a, F_a$.
They generate the infinitesimal symmetry transformations of the
constraint equations:
\begin{eqnarray}&\Delta_a^-\, V_{m,n}=F_a \, V_{m,n}\,,\quad
\Delta_{ab}\, V_{m,n} = K_{ab} \, V_{m,n}&\nonumber\\
&\Delta_a^+\, V_{1,n} = E_a \, V_{1,n}\,,\quad
\Delta_a^+\,V_{0,n}=E_a\,V_{0,n} + i F_a V_{1,n}\,,\quad
\Delta_a^+\, V_{-1,n} = E_a \, V_{-1,n} - i F_a V_{0,n}\,.&\nonumber\\
\label{eq:symce}\end{eqnarray}
We write $E_a$, $F_a$, and $K_{ab}$ for the operators
that represent the left action of $e_a, $ $f_a$, and $t_{ab}$ on
$g$.
In the above coordinates,
\begin{eqnarray}
F_a &=& {d\over d\t^a}\,,\qquad
K_{ab} = -\t_a {d\over d\t^b} + \t_b {d\over d\t^a} + t_{Lab}\nonumber\\
E_a &=& {\textstyle{1\over 2}}\epsilon_{abcd}\,\t^b\,
( t_{L}^{cd} - \t^c {d\over d\t_d}\,) + h_{a\bar b} 
{d\over d\bar\t_{\bar b}}\,,\nonumber\\
\label{eq:efk}\end{eqnarray}
where we have introduced an operator $t_L$
that generates the left action of $SU(2)\times SU(2)$ on $h$ alone,
without acting on the $\theta$'s. Here
\begin{equation}
g = g (x,\t\,, \bar \t)\,
= e^{\t^a f_a}\,
e^{\half\sigma^{p cd} x_p t_{cd}}\, e^{\bar\t^{\bar a} e_{\bar a}}
= e^{\t^a f_a}\, h(x) \, e^{\bar\t^{\bar a} e_{\bar a}},
\end{equation}
\begin{equation} 
t_{Lab} \, g \, = \, e^{\t^a f_a}\, (-t_{ab})
\, h(x) \, e^{\bar\t^{\bar a} e_{\bar a}}\,,
\end{equation}
and we found (\ref{eq:efk}) by requiring
$F_ag=f_ag$, $E_ag=e_ag$,
$K_{ab}g=-t_{ab}g$.
Similar expressions hold for the right-acting generators
$\bar K_{\bar a\bar b}$, $\bar E_{\bar a}$, and $\bar F_{\bar a}$.

The operators $t^{ab}_L, t^{\bar a\bar b}_R$
describe invariant derivatives on the $SO(4)$
group manifold. These can be related to covariant derivatives
${\cal T}_L^{cd} \equiv -\sigma^{p\,cd} \, D_p\,,\, 
{\cal T}_R^{\bar c\bar d} \equiv \sigma^{p\, \bar c\bar d} \, D_p\,\,,$
where for example, acting on a function,
${\cal T}_L=t_L$ and ${\cal T}_R=t_R$. But
when acting on fields that carry vector or spinor indices,
they differ so that for example on spinor indices
$t_L^{ab} V_e=
{\cal T}_L^{ab} \, V_e + \half \delta^a_e\, \delta^{bc} V_c
- \half \delta^b_e\, \delta^{ac} V_c\,.$

In fact, for the Type IIB superstring on
$AdS_3\times S^3\times K3$ with background Ramond flux, 
a sigma model~\cite{bvw}
with conventional local interactions
(no spin fields in the action) was found using the supergroup
$PSU(2|2)$ as target, coupled to ghost fields $\rho$ and $\sigma$.
The spacetime symmetry group is $PSU(2|2)\times PSU(2|2)$,
acting by left and right multiplication on the group manifold,
{\it i.e.}  by $g\to agb^{-1}$ where $g$ is a $PSU(2|2)$-valued field,
and $a,b\in PSU(2|2)$ are the symmetry group's Lie algebra elements.
The supergroup is generated by the super Lie algebra with 12 bosonic
generators forming a subalgebra $SO(4)^2$ together with 16 odd generators.
Hence our model has non-maximal supersymmetry with 16 supercharges.

The $PSU(2|2)$-valued field $g$ is given in terms of $x,\t$, and $\bar\t$,
which are identified as coordinates on the supergroup manifold.
In addition, the Type IIB on $AdS_3 \times S^3\times M$ has worldsheet
variables describing the compactification degrees of freedom on the
four-dimensional space $M$. The vertex operators $V_{mn}(x,\t,\bar\t)$
are examples of the field $g$.

To interpret the generators $E_a, F_a, K_{ab}$, we recall that
in flat space, the $d=6$ supersymmetry algebra for the left-movers is
given by
\begin{eqnarray}
\{ q_a^+, q_c^- \} &=& \half\epsilon_{abcd} P^{cd}\nonumber\\
{[ P_{ab}, P_{cd} ]}  &=& 0
= [ P_{ab} , q_c^\pm ] = \{q_a^+,q_b^+\}=
\{q_a^-, q_b^-\}\nonumber\\
\end{eqnarray}
where $P_{ab} \equiv \delta_{ac}\delta_{bd} P^{cd}$ and
\begin{eqnarray}
q_a^- &=& \oint F_a (z)\nonumber\\
q_a^+ &=& \oint ( e^{-\rho -i\sigma} F_a(z) + i E_a(z) )\label{eq:ssgen}\\
P^{ab} &=& \oint \partial x_m (z) \sigma^{mab}\,.\nonumber\\
\end{eqnarray}
In flat space we have $F_a(z) = p_a(z)$ and
$E_a(z) = \half\e_{abcd} \t^b(z) \partial x_m (z) \sigma^{mcd}$.
We distinguish between the currents and their zero moments
$E_a, F_a$ which together with $P_{ab}$ also generate
the flat space supersymmetry algebra
\begin{eqnarray}
{[ P_{ab}, P_{cd} ]} &=& 0 = [ P_{ab}, F_c ] = [ P_{ab}, E_c ]\,, \nonumber\\
\{ E_a , F_b \} &=& \half \e_{abcd} P^{cd}\,,\qquad
\{ E_a, E_b\} = \{F_a, F_b \} = 0\,.\\
\nonumber\label{eq:fsssa}\end{eqnarray}
On $AdS_3\times S^3$, the Poincare supersymmetry algebra (\ref{eq:fsssa}) 
is replaced by
the $PSU(2|2)$ superalgebra
\begin{eqnarray}
{[ K_{ab}, K_{cd} ] }&=& \delta_{ac} K_{bd} - \delta_{ad} K_{bc} - 
\delta_{bc} K_{ad} + \delta_{bd} K_{ac}\nonumber\\
{[ K_{ab}, E_c ]} &=&  \delta_{ac} E_b - \delta_{bc} E_a\,\qquad
{[ K_{ab}, F_c ]} =  \delta_{ac} F_b - \delta_{bc} F_a\,\qquad\nonumber\\
\{E_a, F_b\} &=& \half e_{abcd} K^{cd}\,\quad
\{E_a, E_b\} = 0 = \{F_a, F_b\} \nonumber\\
\label{eq:adssa}\end{eqnarray}
The generators $q_a^\pm$, which generate the $AdS$ tranformations 
(\ref{eq:symce}),
still have a form similar to (\ref{eq:ssgen}) but
$E_a(z,\bar z), F_a(z,\bar z)$ are no longer holomorphic and their
zero moments with respect to $z$ satisfy (\ref{eq:adssa}).

For the bosonic field components of the vertex operator the $AdS$ constraint
equations (\ref{eq:tone}-\ref{eq:tthree}) result in
\begin{equation}
\Box \,h^g_{\,\,\bar a} \,V^{--}_{ag} =
-4 \,\sigma^m_{ab}\,\sigma^n_{gh}\,\delta^{bh}\, h^g_{\,\,\bar a}\,
G_{mn}
\end{equation}
\begin{equation}
\Box \, h^g_{\,\,\bar a}\,
h^h_{\,\,\bar b}\, \sigma^m_{ab}\,\sigma^n_{gh}\, G_{mn} =
{\textstyle{1\over 4}} \epsilon_{abce} \epsilon_{fghk}
\, \delta^{ch}\,h^f_{\,\,\bar a}\,  h^g_{\,\,\bar b}\,
\, F^{++ e k}
\end{equation}
\begin{equation}
\Box \, h_g^{\,\,\bar a}\,
F^{++ a g} = 0\,,\quad
\Box \, h_g^{\,\,\bar a}\, A_a^{-+  g} = 0\,,\quad
\Box \, h^g_{\,\,\bar a}\, A_g^{+-  a} = 0
\end{equation}
\begin{equation}
\epsilon_{eacd}\, t^{cd}_L\, h^b_{\,\,\bar a}\, A_b^{+-a} = 0\,,\qquad
\epsilon_{\bar e\bar b\bar c\bar d}\, t^{\bar c\bar d}_R\,
h_a^{\,\,\bar a}  \,A_{\bar a}^{-+ \bar b} = 0
\end{equation}
\begin{equation}
\epsilon_{eacd}\, t^{cd}_L\, h_b^{\,\,\bar a}\, F^{++ab} = 0\,,\qquad
\epsilon_{\bar e\bar b\bar c\bar d}\, t^{\bar c\bar d}_R\,
h^a_{\,\,\bar a}  \,F^{++\bar a \bar b} = 0
\end{equation}
\begin{equation}
t^{ab}_L\, h^g_{\,\,\bar a}\,  h^h_{\,\,\bar b}\,\sigma^m_{ab}\,\sigma^n_{gh}\,
G_{mn} =\,0\,,\quad
t^{\bar a\bar b}_R\, h^{\,\,\bar g}_a\,  h^{\,\,\bar h}_{b}\,
\sigma^m_{\bar g\bar h}\,\sigma^n_{\bar a \bar b}\, G_{mn}\,
=\,0\,.
\end{equation}
We have expanded $G_{mn}= g_{mn} + b_{mn} + \bar g_{mn}\phi$.
The $SO(4)$ Laplacian is\break
$\Box\equiv\nobreak{\textstyle{1\over 8}} \epsilon_{abcd} \,t^{ab}_L\, t^{cd}_L
\,=\, {\textstyle{1\over 8}} \epsilon_{\bar a\bar b\bar c\bar d}
\,t^{\bar a\bar b}_R\, t^{\bar c \bar d}_R$.
In order to compare this with the supergravity field theory,
we can use our expressions for the group manifold invarinat derivatives 
terms of covariant derivatives. 
We will also use the fact that
on $AdS_3\times S^3$  we can write the Riemann tensor 
and the metric tensor as 
\begin{eqnarray}
\bar R_{mnp\tau} &=&
{\textstyle{1\over 4}} \, (\, \bar g_{m\tau} \bar R_{np} +
\bar g_{np} \bar R_{m\tau} - \bar g_{n\tau } \bar R_{mp} -
\bar g_{mp} \bar R_{n\tau}\,)\nonumber\\
\bar g_{mn} &=&
{\textstyle{1\over 2}}\, \sigma_m^{ab}\,\sigma_{n\,ab}\,.
\end{eqnarray}
The sigma matrices $\s^{m ab}$ satisfy the algebra
$\s^{m ab} \s^n_{ac} + \s^{n ab} \s^m_{ac}
= \eta^{mn} \delta^b_c$
in flat space, where $\eta^{mn}$ is the six-dimensional Minkowski metric.
Sigma matrices with lowered indices are defined by
$\s^m_{ab} = \half\e_{abcd} \s^{m cd}$, although for other
quantities indices are raised and lowered with $\delta^{ab}$,
so we distinguish $\s^m_{ab}$ from $\delta_{ac}\,\delta_{bd}\,\s^{m cd}$.
In curved space,  $\eta_{mn}$ is  replaced
by the $AdS_3\times S^3$ metric $\bar g_{mn}$.
We then find from the string constraints that the six-dimensional
string field components $g_{mn}, b_{mn},\phi,$ etc. satisfy
(\ref{eq:steqone},\ref{eq:steqtwo}).

\subsection{Correlation Functions}

We will review~\cite{bd} the six-dimensional
three-graviton tree level amplitude in (6d) flat space,
for Type IIB superstrings on ${\bf R^6}\times K3$ 
in the BVW formalism. It is contained in the closed string
three-point function
\be <\,V (z_1,\bar z_1)\,( G^+_0 \bar G^+_0  V (z_2, \bar z_2))
(\tilde G^+_0 \bar{\tilde G}^+_0 V (z_3, \bar z_3)) \,>\,\label{eq:sam}\ee
where the vertex operators are given by 
\be V (z,\bar z) = e^{i\sigma(z) + \rho(z)}\,
e^{i\bar\sigma(\bar z) + \bar\rho(\bar z)}\,
\theta^a(z) \theta^b(z) \,\bar\theta^{\bar a} (\bar z)
\,\bar\theta^{\bar b} (\bar z) \,\sigma_{ab}^m \,\sigma_{\bar a\bar b}^n
\,\phi_{mn}(X(z,\bar z))\,,\label{eq:gravvo}\ee
when the field $$\phi_{mn} = g_{mn} + b_{mn} + \bar g_{mn}\,\phi\,$$
satisfies the constraints we found previously 
$\partial^m\phi_{mn} = 0$, and $\Box \phi_{mn} = 0\,.$
These constraints imply the gauge conditions $\p^m b_{mn} = 0$
for the two-form, and
$\p^m g_{mn} = -\p_n \phi$ for the traceless graviton $g_{mn}$
and dilaton $\phi\,.$
There is a residual gauge symmetry
\be g_{mn}\rightarrow g_{mn} + \p_m\xi_n + \p_n\xi_m \,,
\qquad \phi\rightarrow \phi\,,\qquad b_{mn}\rightarrow
b_{mn}\,\label{eq:residgt}\ee
with $\Box \xi_n = 0\,,\, \p\cdot\xi = 0\,.$
To evaluate (\ref{eq:sam}), we extract the simple poles as
\bea
&G_0^+ \bar G_0^+ \, V (z,\bar z) = e^{i\sigma }\,
e^{i\bar\sigma}\,(-4) \,
\,[ \phi_{mn}(X) \,\partial X^m \bar\partial X^n \cr
&\hskip60pt -p_a \theta^b \,\sigma_{cb}^m\,\sigma^{p c a}
\bar\partial X^n \,\partial_p\,\phi_{mn}(X)\cr
&\hskip60pt -\bar p_{\bar a} \bar\theta^{\bar b} \,\sigma_{\bar c\bar b}^n\,
\sigma^{p \bar c\bar a}
\partial X^m \,\partial_p \phi_{mn}(X)\cr
&\hskip110pt + p_a \theta^b \,\bar p_{\bar a} \bar\theta^{\bar b} \,
\sigma_{cb}^m\,\sigma^{p c a}\,
\sigma_{\bar c\bar b}^n\, \sigma^{q \bar c\bar a}\,
\partial_p\partial_q \,\phi_{mn}(X)\,]\label{eq:gonv}\eea
\be
\tilde G_{0}^+ \bar {\tilde G}_{0}^+
\, V (z,\bar z) = e^{i H_C + 2\rho + i\sigma }\,
e^{i\bar H_C + 2\bar\rho + i\bar\sigma}\,
\theta^a \theta^b \,\bar\theta^{\bar a}
\,\bar\theta^{\bar b} \,\sigma_{ab}^m \,\sigma_{\bar a\bar b}^n
\,\phi_{mn}(X)\,.\label{eq:gtonv}\ee
Using the OPE's for the ghost fields and $H_C$, we partially
compute (\ref{eq:sam}) by evaluating the leading singularities
to find
\vfill\eject
\bea
&\hskip-50pt <\,V_1(z_1,\bar z_1)\,( G^+_0 \bar G^+_0  V_2(z_2, \bar z_2))
(\tilde G^+_0 \bar{\tilde G}^+_0 V_3(z_3, \bar z_3)) \,>\cr
&\hskip-50pt = (z_1-z_2) (z_2-z_3) (z_1-z_3)^{-1}
(\bar z_1 - \bar z_2) (\bar z_2 - \bar z_3) (\bar z_1 - \bar z_3)^{-1}\cr
&\hskip50pt  \cdot 
4 <\, e^{iH_C(z_3)} e^{\rho (z_1) + 2\rho(z_3)} e^{i\sigma (z_1)
+ i \sigma (z_2) + i \sigma (z_3)}\,
e^{i\bar H_C(\bar z_3)} e^{\bar\rho (\bar z_1) + 2\bar\rho(\bar z_3)}
e^{i\bar\sigma (\bar z_1)
+ i \bar\sigma (\bar z_2) + i \bar\sigma (\bar z_3)}\cr
&\hskip-100pt \cdot \, \theta^a(z_1) \theta^b(z_1) \bar\theta^{\bar a}(\bar z_1)
\bar\theta^{\bar b}(\bar z_1) \,\sigma_{ab}^m \,\sigma_{\bar a\bar b}^n
\,\phi_{mn}(X(z_1,\bar z_1))\,\cr
&\hskip-100pt\cdot \,[\,\phi_{jk}(X(z_2,\bar z_2))\,
\partial X^j(z_2) \bar\partial X^k(\bar z_2)\cr
&\hskip-10pt - p_e(z_2) \theta^f(z_2) \,\sigma_{uf}^j\,\sigma^{p u e}
\bar\partial X^k (\bar z_2)\,\partial_p\,\phi_{jk}(X(z_2,\bar z_2))\cr
&\hskip-10pt -\bar p_{\bar e}(\bar z_2)
\bar\theta^{\bar f}(\bar z_2) \,\sigma_{\bar u\bar f}^k\,
\sigma^{p \bar u\bar e}
\partial X^j (z_2)\,\partial_p \phi_{jk}(X(z_2,\bar z_2))\cr
&\hskip60pt + p_e(z_2)\theta^f (z_2)\,
\bar p_{\bar e}(\bar z_2) \bar\theta^{\bar f}(\bar z_2) \,
\sigma_{uf}^j\,\sigma^{p u e}\,
\sigma_{\bar u\bar f}^k\, \sigma^{q \bar u\bar e}\,
\partial_p\partial_q \,\phi_{jk}(X(z_2,\bar z_2))\,]\cr
&\hskip-80pt\cdot\,\theta^c(z_3) \theta^d(z_3) \bar\theta^{\bar c}(\bar z_3)
\bar\theta^{\bar d}(\bar z_3) \,\sigma_{cd}^g \,\sigma_{\bar c\bar d}^h
\,\phi_{gh}(X(z_3,\bar z_3))\,>\,.\eea
Evaluating the remaining $z_2, z_3$ operators products, and
using the $SL(2,C)$ invariance of the amplitude to take the
three points to constants
$z_1\rightarrow\infty$, $\bar z_1\rightarrow\infty$,
$z_2\rightarrow 1, \bar z_2\rightarrow 1$,
$z_3\rightarrow 0, \bar z_3\rightarrow 0$,
we find
\bea
&\hskip-100pt <\,V_1(z_1,\bar z_1)\,( G^+_0 \bar G^+_0  V_2(z_2, \bar z_2))
(\tilde G^+_0 \bar{\tilde G}^+_0 V_3(z_3, \bar z_3)) \,>\cr
&\hskip-100pt 
= (z_2-z_3) (\bar z_2 - \bar z_3) (z_2-z_3)^{-1} (\bar z_2 - \bar z_3)^{-1}
\cdot 4\cr
& \hskip8pt\cdot\,<\, e^{iH_C(0) +3\rho(0) + 3i\sigma(0)}
e^{i\bar H_C(0) + 3\bar\rho(0) + 3i\bar\sigma(0)}
\theta^a_0\theta^b_0\theta^c_0\theta^d_0\,
\bar\theta^{\bar a}_0
\bar\theta^{\bar b}_0\bar\theta^{\bar c}_0\bar\theta^{\bar d}_0\,>\cr
&\hskip-10pt  \cdot\,[ \sigma_{ab}^m \,\sigma_{cd}^g \,
\sigma_{\bar a\bar b}^n\, \sigma_{\bar c\bar d}^h
\, \,<\phi_{mn}(X(\infty))\,
\phi_{jk}(X(1)) \,\partial^j\partial^k\phi_{gh}(X(0))>\cr
&\hskip30pt  +2 \sigma_{ab}^m \, (\sigma^j\sigma^p\sigma^g)_{cd}\,
\sigma_{\bar a\bar b}^n\, \sigma_{\bar c\bar d}^h\,
\,<\phi_{mn}(X(\infty))\,
\partial_p \phi_{jk}(X(1))
\, \partial^k\phi_{gh}(X(0))>\cr
&\hskip30pt +2 \sigma_{ab}^m \, \sigma^g_{cd}\,
\sigma_{\bar a\bar b}^n\, (\sigma^k\sigma^p\sigma^h)_{\bar c\bar d}\,
\,<\phi_{mn}(X(\infty))\,
\partial_p \phi_{jk}(X(1))
\, \partial^j\phi_{gh}(X(0))>\cr
&+4 \sigma_{ab}^m \, (\sigma^j\sigma^p\sigma^g)_{cd}\,
\sigma_{\bar a\bar b}^n\, (\sigma^k\sigma^q\sigma^h)_{\bar c\bar d}\,
\,<\phi_{mn}(X(\infty))\,
\partial_p\partial_q \phi_{jk}(X(1))
\,\phi_{gh}(X(0))>]\cr
&\eea
which results in 
\bea
&\hskip-100pt =4\,[\,\,\bar g^{mg}\bar g^{nh}
\, \,<\phi_{mn}(x_0)\,
\phi_{jk}(x_0) \,\partial^j\partial^k\phi_{gh}(x_0)>\cr
&\hskip17pt - \bar g^{nh} \,
(\sigma^m\sigma^j\sigma^p\sigma^g)^d_{\hskip2pt d}\,\,<\phi_{mn}(x_0)\,
\partial_p \phi_{jk}(x_0)
\, \partial^k\phi_{gh}(x_0)>\cr
&\hskip17pt - \bar g^{mg}\,
(\sigma^n\sigma^k\sigma^p\sigma^h)^{\bar d}_{\hskip2pt\bar d}\,
\,<\phi_{mn}(x_0)\,
\partial_p \phi_{jk}(x_0)
\, \partial^j\phi_{gh}(x_0)>\cr
&\hskip17pt + (\sigma^m\sigma^j\sigma^p\sigma^g)^d_{\hskip2pt d}\,\,
(\sigma^n\sigma^k\sigma^q\sigma^h)^{\bar d}_{\hskip2pt\bar d}\,
\,<\phi_{mn}(x_0)\,
\partial_p\partial_q \phi_{jk}(x_0)
\,\phi_{gh}(x_0)>]\,.\cr
&\eea
The second equality follows from the vacuum expectation value
of the ghost fields, $H_C$ and eight fermion zero modes
\be <\, e^{iH_C(0) +3\rho(0) + 3i\sigma(0)}
e^{i\bar H_C(0) + 3\bar\rho(0) + 3i\bar\sigma(0)}
\theta^a_0\theta^b_0\theta^c_0\theta^d_0\,
\bar\theta^{\bar a}_0
\bar\theta^{\bar b}_0\bar\theta^{\bar c}_0\bar\theta^{\bar d}_0\,>
 = {\textstyle{1\over 16}}
\epsilon^{abcd}\,\epsilon^{\bar a\bar b\bar c\bar d}\,.\ee
We have also used various sigma matrix identities.
Since $(\sigma^m\sigma^n\sigma^p\sigma^q)^d_{\hskip2pt d}
= \bar g^{mn}\bar g^{pq} + \bar g^{mq}\bar g^{np}
- \bar g^{mp}\bar g^{nq}$ where in flat space $\bar g_{mn} = \eta_{mn}$,
and using the gauge condition 
$\partial^m\phi_{mn} = 0$ once more,
we find  
\bea
&\hskip-50pt <\,V_1(z_1,\bar z_1)\,( G^+_0 \bar G^+_0  V_2(z_2, \bar z_2))
(\tilde G^+_0 \bar{\tilde G}^+_0 V_3(z_3, \bar z_3)) \,>\cr
&\hskip-50pt =4\,[\,\,\bar g^{mg}\bar g^{nh}
\, \,<\phi_{mn}(x_0)\,
\phi_{jk}(x_0) \,\partial^j\partial^k\phi_{gh}(x_0)>\cr
&\hskip50pt - \bar g^{nh} (\bar g^{mj}\bar g^{pg}
- \bar g^{mp}\bar g^{jg} )\,
\,<\phi_{mn}(x_0)\,
\partial_p \phi_{jk}(x_0)
\, \partial^k\phi_{gh}(x_0)>\cr
&\hskip50pt - \bar g^{mg}\,
(\bar g^{nk}\bar g^{ph} - \bar g^{np}\bar g^{kh} )\,
\,<\phi_{mn}(x_0)\,
\partial_p \phi_{jk}(x_0)
\, \partial^j\phi_{gh}(x_0)>\cr
&\hskip-30pt + (\bar g^{mj}\bar g^{pg}
- \bar g^{mp}\bar g^{jg} )\,
(\bar g^{nk}\bar g^{qh} - \bar g^{nq}\bar g^{kh} ))\cr
&\hskip10pt \cdot <\phi_{mn}(x_0)\,
\partial_p\partial_q \phi_{jk}(x_0)
\,\phi_{gh}(x_0)>]\cr
&= 12\,[\, <\phi^{mn}(x_0)\,\phi^{jk}(x_0)\,
\partial_m\partial_n \phi_{jk}(x_0)>
+ 2\, <\phi^{mn}(x_0)\,\partial_m\phi^{jk}(x_0)\,
\partial_j \phi_{nk}(x_0)> \,]\,.\cr
&\label{eq:thrgrav}\eea
To compare this with the supergravity field theory,
we consider the 
Einstein-Hilbert action 
$I = \int d^dx {\sqrt{|g|}} \{-{R\over{2 \kappa^2}}\}\,.$
Expanding to third order in $\kappa$ using
$g_{\mu\nu} = \eta_{\mu\nu} + 2 \kappa h_{\mu\nu}$, we find
the three-point interaction $I_3$.
In harmonic gauge, {\it i.e.} when
$\p^\mu h_{\mu\nu} - \half \p_\nu h^\rho_\rho = 0$, and
on shell $\Box h_{\mu\nu} = 0$, the cubic coupling is given by
\be
I_3 = - \kappa \int d^dx [ h^{\mu\nu} h^{\rho\sigma} \partial_\mu\partial_\nu
h_{\rho\sigma} + 2 h^{\mu\nu} \partial_\mu h^{\rho\sigma} \partial_\rho
h_{\nu\sigma} ]\,.\label{eq:flat}\ee
The gauge transformations
\be
h_{\mu\nu} \rightarrow h_{\mu\nu} + \p_\mu\xi_\nu
+ \p_\nu\xi_\mu\label{eq:gtran}\ee
leave invariant the harmonic gauge condition and
$I_3$, given in (\ref{eq:flat}) , when $\Box\xi_\mu = 0\,.$
With this gauge symmetry, we could further choose
$h^\rho_\rho = 0$, $\p^\mu h_{\mu\nu} = 0$.
Then $I_3$ is the three-graviton amplitude, and
it is invariant under residual gauge transformations
that have $\p\cdot\xi = 0$.

To identify the string theory three-graviton amplitude
from (\ref{eq:thrgrav}), we set $b_{mn}$ to zero, and
use the field identifiations that relate
the string fields $g_{mn}, \phi$ to the supergravity
field $h_{mn}$ via $\phi\equiv -{\textstyle{1\over 3}}h^\rho_\rho$
and $g_{mn}\equiv h_{mn} - {\textstyle{1\over 6}} \bar g_{mn}
h^\rho_\rho$, where $h_{mn}$ is in harmonic gauge.
Then $\phi_{mn} = h_{mn} -\half\bar g_{mn}
h^\rho_\rho\,,$ and from (\ref{eq:thrgrav})
the on shell string tree amplitude is
\bea
&\hskip-100pt {\textstyle{-{\kappa\over {12}}}}
<\,V_1(z_1,\bar z_1)\,( G^+_0 \bar G^+_0  V_2(z_2, \bar z_2))
(\tilde G^+_0 \bar{\tilde G}^+_0 V_3(z_3, \bar z_3)) \,>\cr
&\hskip-30pt = -\kappa \int d^d x \, [ \, \phi^{mn}(x)\,\phi^{jk}(x)\,
\partial_m\partial_n \phi_{jk}(x)
+ 2\, \phi^{mn}(x)\,\partial_m\phi^{jk}(x)\,
\partial_j \phi_{nk}(x) \,]\cr
&=  - K \int d^d x \, [ h^{mn} h^{jk}\, \p_m \p_n h_{jk}\,
+ 2\, h^{mn} \p_m h^{jk} \p_j h_{nk}\,]
\,+ \kappa  \int d^d x \, h^{mn} \,\p_m h^k_k \,\,\p_n h^p_p  \cr
&\hskip-300pt = I_3  + I'_3\label{eq:thrgravft}\eea
where $I'_3$ is the one graviton - two dilaton amplitude,
$I_3$ is the three graviton interaction in harmonic gauge,
and $d=6$.
$I_3$ and $I'_3$ are invariant separately
under the gauge transformation (\ref{eq:gtran}) with
$\Box \xi_n = 0$ and $\p\cdot\xi = 0$, which corresponds to
the gauge symmetry of the string field
$\phi_{mn}\rightarrow \phi_{mn} + \p_m\xi_n + \p_n\xi_m\,.$
$I_3$ by itself is also invariant under gauge transformations
for which $\p\cdot\xi\ne 0$, and these can be used to
eliminate the trace of $h_{mn}$ in $I_3$.
In the string gauge, the trace of $\phi_{mn}$ is related to
the dilaton $\phi^m_m =  6\phi $, so even when $b_{mn} = 0$, 
(\ref{eq:thrgrav})
contains both the three graviton amplitude and the one graviton - two
dilaton interaction.
So it turns out we could have extracted $I_3$ from (\ref{eq:thrgrav}) 
simply by
setting both $b_{mn}=0$ and $\phi = 0$, since then $\phi_{mn} = g_{mn}$
and $\p^m g_{mn} = 0\,.$

Correlation functions on $AdS_3\times S^3$ have also been studied~\cite{bd}.

\section{Concluding Remarks}

Type IIB superstrings on $AdS_3\times S^3\times K3$
can have either Neveu-Schwarz {\it or} Ramond background
flux to ensure the background metric is a solution to the 
equations of motion. The Neveu-Schwarz case corresponds to
a $WZW$ model and has been extensively studied~{\cite{tone}-\cite{mothree}}.
Since these two cases are $S$-dual to each other, 
the massless spectrum is the same, but the perturbative massive
spectrum will be different. For the Type IIB superstring on
$AdS_5\times S^5$, the flux supporting the metric can only be Ramond
~{\cite{kvnr}-\cite{gb}.
Thus the $AdS_3\times S^3$ analysis discussed in these lectures
is meant as a step towards the $AdS_5$ case. 

The conjectured duality between M-theory or Type IIB string theory
on anti-de Sitter (AdS) space and the conformal field theory on
the boundary of AdS space~{\cite{ffz}-\cite{gub}} 
may be useful in giving a controlled
systematic approximation for strongly coupled gauge theories.
The formulation of vertex operators and
string theory tree amplitudes for the IIB superstring on $AdS_5\times S^5$
will allow access
to the dual conformal $SU(N)$ gauge field theory $CFT_4$ at large $N$, but
{\it small} fixed `t Hooft coupling $x = g^2_{YM} N$ in the dual
correspondence, as
$(g^2_{YM} N)^\half (4\pi)^\half = R^2_{sph}/{\alpha'}$.
Presently only the large $N$, and {\it large} fixed `t Hooft coupling $x$
limit is accessible in the $CFT$, since only the supergravity limit
($\alpha'\rightarrow 0$) of the correlation functions
of the AdS theory is known.

Tree level $n$-point correlation functions for $n\ge 4$
presumably have $\alpha'$ corrections, since the worldsheet theory is
not a free conformal field theory. However, 
there may be sufficiently many
symmetry currents to determine the tree level
correlation functions exactly in $\alpha'$ as well.
This might be possible via integrable methods for
sigma models which have a supergroup manifold target space~\cite{bzv,rs} 
such as
the $AdS_3\times S^3$ theory.

\section*{Acknowledgments}

I would like to thank the organizers of TASI '01 for the invitation 
to lecture on
these topics and for providing a very lively and hospitable atmosphere. 
This work is partially supported by the U.S. Department of Energy under
Grant No. DE-FG02-97ER-41036/Task A.


\end{document}